\documentclass[reprint,amsmath,amssymb,aps,prd,floatfix,nofootinbib]{revtex4-2}

\usepackage{graphicx}
\usepackage{subfigure}
\usepackage{dcolumn}
\usepackage{bm}
\usepackage{booktabs}
\usepackage{color,verbatim}
\usepackage{braket,dsfont}
\usepackage{physics, wasysym }
\usepackage[colorlinks=true,citecolor=blue,filecolor=blue,linkcolor=blue,urlcolor=blue,pdftex]{hyperref}
\usepackage{gensymb}
\usepackage[normalem]{ulem}

\newcommand{\mnras}{MNRAS}
\newcommand{\aap}{A\&A}
\newcommand{\apjs}{ApJS}
\newcommand{\jcap}{JCAP}

\begin{document}

\preprint{APS/123-QED}

\title{Searching for dilaton fields in the Ly$\alpha$ forest}

\author{Louis Hamaide}
\email{louis.hamaide@kcl.ac.uk}
\affiliation{
King’s College London, Strand, London, WC2R 2LS, UK}
 
\author{Hendrik M\"uller}
\altaffiliation{Both first authors have contributed equally to this work.}
\affiliation{Max-Planck-Institut für Radioastronomie, Auf dem Hügel 69, D-53121 Bonn (Endenich), Germany
}

\author{David J. E. Marsh}
\affiliation{King’s College London, Strand, London, WC2R 2LS, UK}

\date{\today}

\begin{abstract}
Dilatons (and moduli) couple to the masses and coupling constants of ordinary matter, and these quantities are fixed by the local value of the dilaton field. If, in addition, the dilaton with mass $m_\phi$ contributes to the cosmic dark matter density, then such quantities oscillate in time at the dilaton Compton frequency. We show how these oscillations lead to broadening and shifting of the Voigt profile of the Ly$\alpha$ forest, in a manner that is correlated with the local dark matter density. We further show how tomographic methods allow the effect to be reconstructed by observing the Ly$\alpha$ forest spectrum of distant quasars. We then simulate a large number of quasar lines of sight using the lognormal density field, and forecast the ability of future astronomical surveys to measure this effect. We find that in the ultra low mass range $10^{-32}\text{ eV}\leq m_\phi\leq 10^{-28}\text{ eV}$ upcoming observations can improve over existing limits to the dilaton electron mass and fine structure constant couplings set by fifth force searches by up to five orders of magnitude. Our projected limits apply assuming that the ultralight dilaton makes up a few percent of the dark matter density, consistent with upper limits set by the cosmic microwave background anisotropies.
\end{abstract}

\maketitle

\section{\label{sec: intro}Introduction}
Dilatons and moduli (including the volume modulus, and radions) are scalar degrees of freedom of string theory and other extra-dimensional theories, which arise in the low energy effective theory after compactification \cite{BeckerBeckerSchwarzBook,Arvanitaki:2014faa}. These fields appear in the scalar potential, which can in some cases lead to their having extremely small masses. Couplings to the Standard Model arise in a variety of ways. Moduli, for example, appear in the gauge kinetic function, with the scalar moduli giving the value of the fine structure constant (pseudoscalar axions fix the Chern-Simons term). The dilaton itself couples to all fields via the Einstein frame metric. For brevity in what follows we refer to such fields collectively as dilatons.

More generally, since string theory contains no dimensionful constants, all the properties of low energy physics must be determined by the values of (scalar) fields. The observed low energy ``constants'' therefore only appear so, with the values fixed only due to the field taking a particular location in some local minimum of the scalar potential. The fields would generically be displaced from this minimum at early times (due to e.g. thermal or quantum fluctuations). Motion from the initial location to the local minimum results in damped oscillations about the minimum. If such a field, $\phi$, is furthermore stable on cosmological time scales, then today the relic oscillations behave as a contribution to the dark matter density of the Universe~\cite{Turner:1983he}. If initial displacements of $\phi$ from the vacuum are of order of the GUT scale, the correct relic abundance is achieved for masses $m_\phi\approx 10^{-20}\text{ eV}$.~\footnote{For heavy, unstable particles, the corresponding phenomena result in the ``cosmological moduli problem''~\cite{Coughlan:1983ci}, restricting unstable moduli to be heavier than around 100 TeV.} In such a scenario, the constants of nature oscillate with a frequency given by the dark matter mass, and an amplitude related to the local dark matter density. A number of surveys have already searched for such effects for dark matter fully composed of dilatons, unsuccessfully (for a review see e.g. \cite{Safronova:2017xyt}, or novel ideas in \cite{Berengut:2010ht}). However, we are encouraged by Webb \textit{et al.}'s \cite{King:2012id,Wilczynska:2020rxx} searches for dipole variations in $\alpha$ on cosmological scales, as this could caused by a $m_\phi\approx\mathcal{O}(10^{-32})$~eV dilaton. In the following, we show how to search for higher mode oscillations of $\alpha$ using the Ly$\alpha$ forest and consequently probe higher dilaton masses.

The Ly$\alpha$ forest is a prominent absorption feature in the spectra of distant quasars bluewards of the Ly$\alpha$ absorption line. It consists of densely packed, narrow absorption lines caused by the absorption of quasar light by intervening neutral hydrogen in the intergalactic medium (IGM) along the line of sight \cite{Bahcall1965, Gunn1965, Bi1997}. The optical depth of the Ly$\alpha$ absorption in a single absorption profile is proportional to the column density of neutral hydrogen. Hence, the Ly$\alpha$ forest is an excellent tomographic tracer for cosmic large scale structures \cite{Bi1992, Hui1997}. Many studies targeted the line profile of absorption lines in the Ly$\alpha$ forest to study IGM physics, e.g. by using its curvature \cite{Becker2011, Boera2014, Gaikwad2021}, by a wavelet analysis \cite{Lidz2010, Garzilli2012, Gaikwad2021, Wolfson2021}, with the 1D-flux power spectrum \cite{Boera2019, Gaikwad2021} and the distribution of Doppler parameters \cite{Rudie2012, Bolton2014, Gaikwad2021}. Recently we developed novel methods \cite{Mueller2020} and software \cite{reglyman} to compute the direct deconvolution of the neutral hydrogen fluctuations from the Voigt-profile for highest resolution ($R\gtrsim50000$) spectra and applied this procedure to UVES SQUAD data \cite{Murphy2019} to obtain measurements of the IGM temperature both consistent with and more accurate then existing methods \cite{Mueller2021}. 
\begin{figure*}[t!]
\includegraphics[width=1.8\columnwidth]{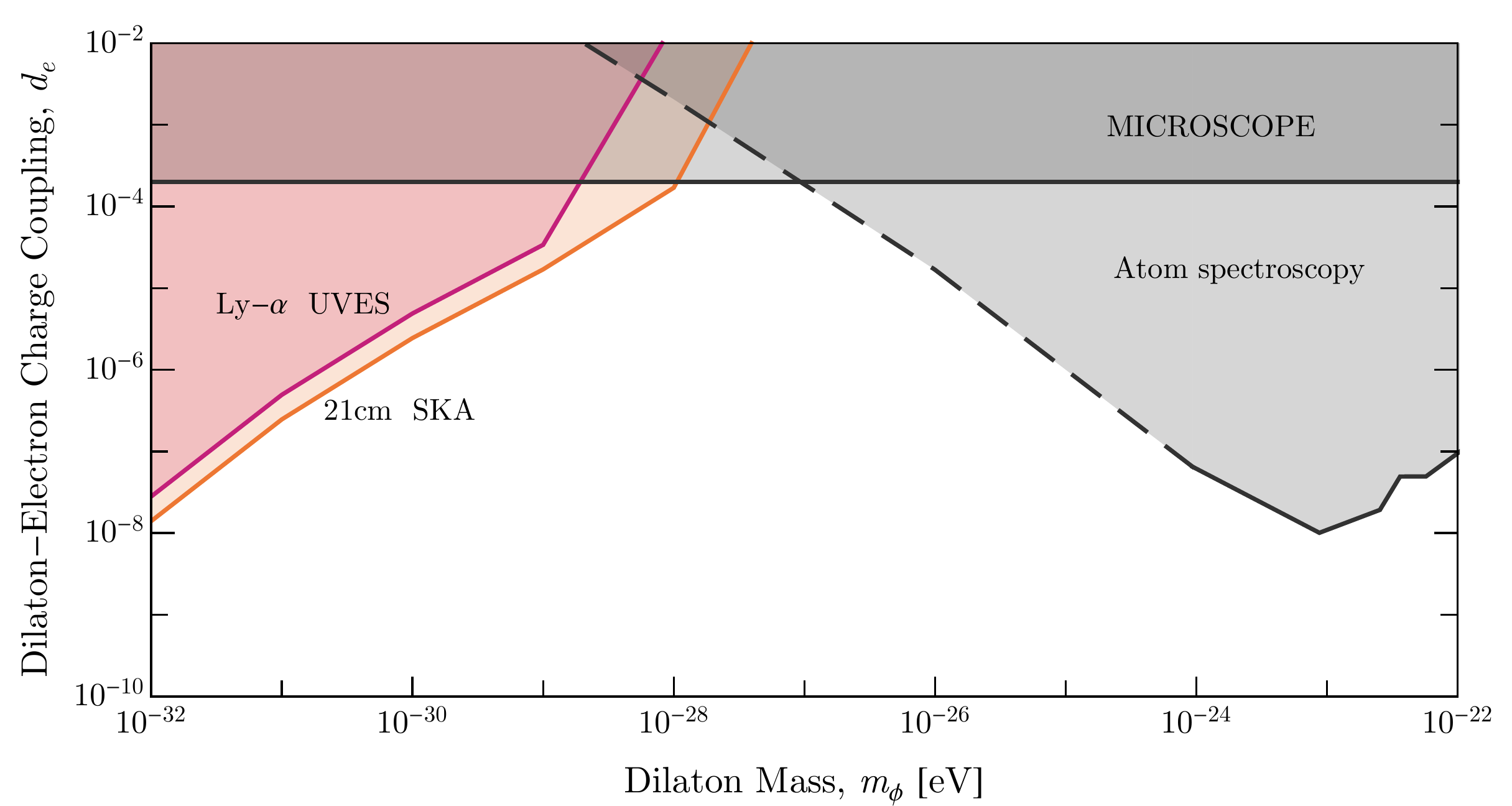}
\caption{\label{fig:de_constraints}
Projected 90\%C.L. constraints on the dilaton coupling $d_e$ as a function of dilaton mass $m_\phi$ possible with Ly$\alpha$ forest spectra. The projected constraints are derived for mock UVES SQUAD Ly$\alpha$ data (red) and adapted for a 21 cm HI survey with SKA-like imaging capabilities (orange). Note we assume dilaton DM fractions given in Table~\ref{tab: implementations}, consistent with measurements of the CMB and matter power spectra. Existing laboratory constraints from the MICROSCOPE fifth force search~\cite{Berge:2017ovy} and atomic spectroscopy~\cite{Hees:2016gop} are shown in dark grey. Dashed lines indicate extrapolation of \cite{Hees:2016gop} for $m_\phi\lesssim 10^{-24}$~eV, adjusted for the maximum allowed dilaton abundance in each mass bin (see Table~\ref{tab: implementations}).
}
\end{figure*}

As we will show, the oscillation of the fine-structure constant induced by dilaton DM affects the Ly$\alpha$ forest by shifting the wavelength of the Ly$\alpha$ transition. For larger dilaton masses ($m_\phi \gtrsim 10^{-28}\,\mathrm{eV}$) the dilaton undergoes several oscillations while a photon travels through an overdensity in the IGM. In this case oscillation of $\alpha$ or $m_e$ appears as an additional broadening of the absorption line, similar to thermal broadening. However, for the smallest masses ($m_\phi \lesssim 10^{-28}\,\mathrm{eV}$), i.e. the longest oscillation times, only partial oscillation occurs while a photon passes through an overdensity, leading to a systematic shift of the absorption lines in redshift space. Both effects modify the absorption profiles along the line of sight in the Ly$\alpha$ forest, and hence may be detectable in high resolution spectra of quasars, using the tools we developed in Refs.~\cite{Mueller2020, Mueller2021}.

\begin{figure*}[t!]
\includegraphics[width=1.8\columnwidth]{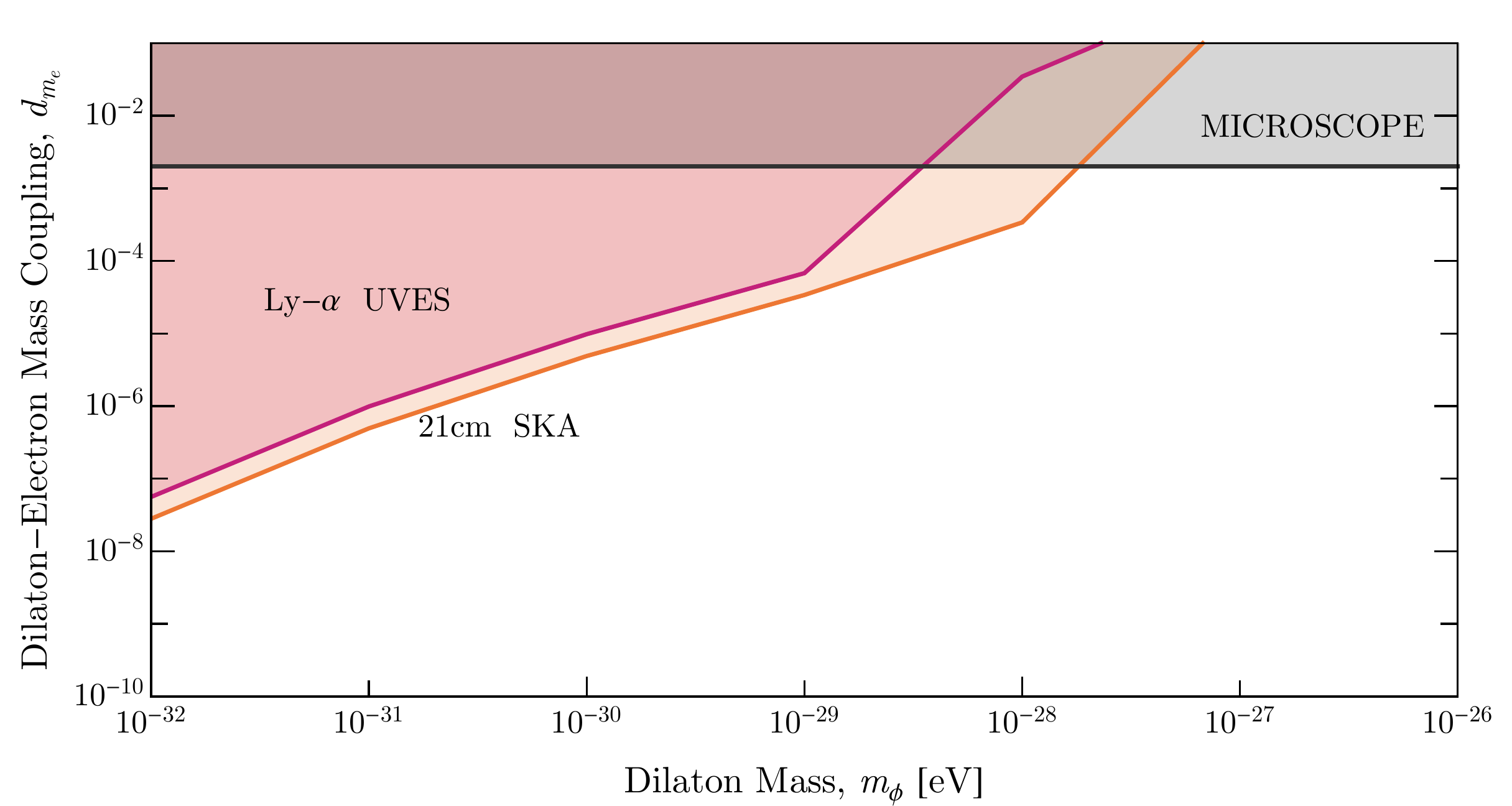}
\caption{\label{fig:dm_constraints}
Projected and existing constraints on the dilaton coupling $d_e$ as a function of dilaton mass $m_\phi$, in the 90\% confidence limit. The projected constraints are from analyzing mock Ly$\alpha$ data (red) and adapted for an SKA-like telescope's imaging capabilities (orange). Competing constraints are from MICROSCOPE \cite{Berge:2017ovy}.
}
\end{figure*}

Following this we create synthetic Ly$\alpha$ forest spectra for a wide range of dilaton masses (ranging from $m_\phi = 10^{-20}\,\mathrm{eV}$ to $m_\phi = 10^{-32}\,\mathrm{eV}$) and compute mock constraints and forecasts for upcoming surveys. Our method gives the strongest constraints for smallest masses, and for the surveys UVES and SKA could significantly outperform ``fifth force'' constraints such as Ref.~\cite{Berge:2017ovy}. Our main results concerning the dilaton coupling to $\alpha$ and $m_e$ are summarized in Figs.~\ref{fig:de_constraints} \& \ref{fig:dm_constraints} respectively.

The paper is structured as follows: in Sec. \ref{sec: the} we present the theoretical basics and discuss the effects of dilatons on the Ly$\alpha$ forest. In Sec. \ref{sec: software} we present our software and synthetic data set. We present our results in Sec. \ref{sec: results} and discuss possible extensions, future directions and drawbacks in \ref{sec: discussion}.

For the rest of the manuscript we use Planck 18 cosmology \cite{Planck:2018vyg}.

\section{\label{sec: the} Theory}

\subsection{\label{sec: dilaton} Dilaton Dark Matter}

The dilaton couplings to the Standard Model arise from the action (we adopt some of the notation of \cite{Damour:2010rp,Arvanitaki:2014faa} in the Einstein frame):
\begin{align} \nonumber
   \mathcal{S} =\int dx^4 & \sqrt{\abs{g}} \left( \frac{1}{2}\partial_{\mu}\phi \partial^{\mu}\phi -V(\phi) + \mathcal{L}_\mathrm{SM} + \mathcal{L}_{\phi ,int} \right)
   \\
   \mathcal{L}_{\phi ,int} & =\frac{\sqrt{4\pi}}{M_\mathrm{Pl}}\phi\left( \frac{d_e}{4e^2}F_{\mu\nu}F^{\mu\nu} - d_{m_e}m_ee\bar{e}
   \right)
\end{align}
where $\abs{g}$ is the determinant of the metric, $\mathcal{L}_\mathrm{SM}$ and $\mathcal{L}_\mathrm{\phi ,int}$ are the Standard Model and dilaton Lagrangians respectively, $F_{\mu\nu}$ is the electromagnetic field tensor, $\bar{e}=\gamma_0 e$ where $e$ is an electron spinor wavefunction, and $\phi$ is the canonically normalised dilaton field from a local change of variables. Except when introducing Ly$\alpha$ forest and the Voigt profile in Secs. \ref{ssec: ly-alpha-forest} and \ref{ssec: mod-voigt-profile}, we use units where $\hbar=c=1$. We take the dilaton potential to be $V(\phi)=\frac{1}{2}m_\phi^2\phi^2$, i.e. a simple mass term, valid for small displacements from the vacuum. Implications of more general potentials have been explored in \cite{Marsh:2011gr,Arvanitaki:2014faa}. We have purposefully omitted other linear scalar couplings such as kinetic gauge field terms of the form $\phi G_{\mu\nu}^AG_{\mu\nu}^A$, quark mass terms $\phi\psi_i\dagger\psi_i$, or a Higgs portal term $\phi H^\dagger H$ \cite{Damour:1994zq,Badurina:2019hst,Arvanitaki:2014faa} and quadratic (or higher order) terms. This is because we are interested in looking for changes in the energies of \emph{atomic} states. We mention here that, as discussed in \cite{Arvanitaki:2014faa} and references therein, in the presence of CP violating supersymmetric physics axions and axion-like particles can couple to an electron mass term as well.

The above couplings $(d_e,d_{m_e})$ can be absorbed into a rescaled electromagnetic and electron mass terms such that:
\begin{align} \nonumber
    \mathcal{L}_\mathrm{EM} = - & \frac{1 -d_e\kappa\phi}{4e^2}F_{\mu\nu}F^{\mu\nu} \approx - \frac{1}{4(1+d_e\kappa\phi)e^2} F_{\mu\nu}F^{\mu\nu}
    \\
    & \alpha \rightarrow \alpha+\delta\alpha = \alpha(1+d_e\kappa\phi) & \label{eq: fine-structure}
\end{align}

and 
\begin{align}
    m_e\bar{e}e \rightarrow m_e(1+d_{m_e}\kappa\phi)\bar{e}e
\end{align}
where $\kappa=\frac{\sqrt{4\pi}}{M_\mathrm{Pl}}$, $\alpha=e^2/4\pi$ and $e^2$ here (and in rest of this paper) refers to the squared electric charge and $\bar{e}e$ refers to the norm of the electron spinor. We thus see that the local value of the dilaton field determines the local observed value of the fine structure constant. Exchange of virtual dilaton particles also mediates new Yukawa forces between Standard Model particles, which we discuss briefly in Sec. \ref{sec: results}.

A local displacement from the vacuum expectation value of the dilaton field arises if all or some of the observed cosmic DM abundance, $\Omega_{\mathrm{DM}} h^2 = \bar{\rho}_{\mathrm{DM}}/(8.07 \times 10^{-11}$~eV$^4)$~\cite{Planck:2018vyg} (where overbar denotes spatial average, and $h$ is the reduced Hubble rate, $H_0=100 h\,\text{ km s}^{-1}\text{ Mpc}^{-1}$), is composed of dilatons. The large occupation number of DM particles throughout the Universe (we consider cases where the dilaton composes more than 1\% of the total DM) permits a description in terms of a classical field. 

Dilaton DM can be produced in the early Universe by the misalignment mechanism, similarly to the well known case of axion and scalar field DM~\cite{Abbott:1982af,Preskill:1982cy,Dine:1982ah,Turner:1983he}. The background homogeneous field evolves according to the Klein-Gordon equation:
\begin{equation}
    \ddot{\phi}+3H\dot{\phi}+m_\phi^2\phi = 0\, ,
    \label{eqn:klein-gordon}
\end{equation}
where $H = \dot{a}/a$ is the Hubble parameter and $a$ the cosmic scale factor, and dots denote derivatives with respect to cosmic time, $t$. The Hubble parameter is determined by the Friedmann equation:
\begin{equation}
    H^2 = \frac{8\pi G_N}{3}\rho\, ,
    \label{eqn:friedmann}
\end{equation}
with $\rho$ the energy density. We assume a standard $\Lambda$CDM cosmology to fix $\rho$, containing radiation, baryons, DM, and the cosmological constant. 

With initial condition $\phi(t_i)=\phi_i$ and $\dot{\phi}(t_i)\approx 0$, the energy density in $\phi$ today ($\Omega_\phi$) is found by solution of Eqs.~(\ref{eqn:klein-gordon},\ref{eqn:friedmann}). Such an initial displacement is expected to be generated, for example, during inflation, and follows in any theory where, in accordance with observation, the initial state of the hot big bang phase is not the vacuum. The Hubble term in Eq.~\eqref{eqn:klein-gordon} acts as a friction, preventing $\phi$ from moving to the vacuum until such a time as $H\lesssim m_\phi$, after which $\phi$ undergoes damped oscillations. At late times, the solution is approximated by $\phi\propto a^{-3/2}\cos m_\phi t$. The energy density scales as $\rho_\phi\propto a^{-3}$ when $H\ll m_\phi$, thus leading to a relic density of dilaton DM (for approximate analytic formulae, see Ref.~\cite{Marsh:2015xka}).

\subsection{Structure Formation}

Structure formation with dilaton DM proceeds as for standard $\Lambda$CDM via gravitational instability from initial, approximately scale invariant, curvature perturbations in the primordial plasma~\cite{Peebles:1982ff}. The curvature perturbations seed initial fluctuations in the modes, $\delta \phi_k$, of the dilaton field on all scales, and in the ``growing mode'', such that $\delta\dot{\phi}_k>0$ (detailed solutions can be found in Ref.~\cite{Hlozek:2014lca}). When $H<m_\phi$, all $\delta\phi_k$ modes begin to oscillate. The evolution can be approximated as:
\begin{equation}
    \delta\phi_k = \psi_k e^{im_\phi t}+\psi^*_k e^{-im_\phi t}\, ,
\end{equation}
where $\psi_k$ is a slowly evolving function of $t$, i.e. $\dot{\psi}\ll m_\phi \psi$. In the non-relativistic limit, the density of DM in mode $k$ is given by:
\begin{equation}
    \rho_k = \frac{1}{2}m_\phi^2|\psi_k|^2\, ,
\end{equation}

Taking $m_k\psi_k = \sqrt{\rho_k} e^{i\theta_k}$ (the ``Madelung form''), the Klein-Gordon equation can be reduced to fluid equations for the dilaton overdensity, $\delta_{\phi,k}=(\rho_k-\bar{\rho}_\phi)/\bar{\rho}_\phi$, and velocity field $v_k=ik \theta_k$. In the non-relativistic limit, the dilaton fluid has an effective sound speed (see e.g. \cite{Bauer:2020zsj}):
\begin{equation}
    c_s^2 = \frac{k^2}{4m_\phi^2 a^2}\, ; \quad (k\ll 2m_\phi a)\, .
    \label{eqn:sound_speed}
\end{equation}
The initial perturbations in the dilaton field begin to grow significantly in the matter dominated era, $z\lesssim 3400$ (where $z$ is the cosmic redshift). The fluctuations are described by the matter power spectrum, $P(k)$. The speed of sound, Eq.~\eqref{eqn:sound_speed}, leads to some modes with small $k$ behaving as cold (collisionless, pressureless) DM, with standard linear growth of fluctuations. Small scale (large $k$) modes, on the other hand, oscillate rather than grow. The scale of separation between growing and oscillating modes is called the Jeans scale~\cite{Khlopov:1985jw}, and can be thought of as the cosmic de Broglie wavelength~\cite{Hlozek:2014lca}.

The Jeans scale can be found analytically for a Universe dominated by dilaton DM, and is given by (e.g. Refs.~\cite{Hu:2000ke,Marsh:2015xka}):
\begin{equation}
    k_J = 66.5 a^{1/4}\left(\frac{\Omega_\phi h^2}{0.12}\right)^{1/4}\left( \frac{m_\phi}{10^{-22}\text{ eV}}\right)^{1/2}\text{ Mpc}^{-1}\, .
\end{equation}
The presence of the Jeans scale causes $P(k)$ to be suppressed in models with a component of dilaton DM compared to pure CDM. This is because dilaton modes with $k>k_J$ experience less growth than those with $k<k_J$. Moreover, the power spectrum contains damped oscillations at large $k$~\cite{Amendola:2005ad,Arvanitaki:2009fg,Marsh:2010wq}.

The power spectrum $P(k)$ is used to place constraints on DM composed entirely of dilatons (for a compilation of $P(k)$ measurements, see Refs.~\cite{Tegmark:2002cy,Chabanier:2019eai,Sabti:2021unj}). The strongest constraint on pure dilaton DM is derived from the Ly$\alpha$ forest flux power spectrum, which demands $m_\phi>2\times 10^{-20}\text{ eV}$ at 95\% credibility~\cite{Rogers:2020ltq}. A weaker, but independent limit can be found using the weak lensing galaxy shear correlation function \cite{Dentler:2021zij}. A standard ``rule of thumb'' limit is $m_\phi\gtrsim 10^{-22}\text{ eV}$, which is confirmed by a variety of measurements, including high redshift galaxy formation~\cite{Bozek:2014uqa,Schive:2015kza,Corasaniti:2016epp}.

For $m_\phi<2\times 10^{-20}\text{ eV}$, dilaton DM is permitted to compose only a fraction of the total observed DM abundance, $\Omega_\phi/\Omega_d<1$. Our constraints on this scenario are derived from cosmological observations of the CMB power spectrum~\cite{Hlozek:2014lca}. The CMB lensing, galaxy power spectrum, and Ly$\alpha$ forest flux power spectrum can also be used to exclude the existence of steps in $P(k)$~\cite{Amendola:2005ad,Hlozek:2017zzf,Lague2021,Kobayashi:2017jcf}. The constraints on $\Omega_\phi/\Omega_d$ from this data is summarized in Table~\ref{tab: implementations}, in dilaton mass bins ranging from $10^{-20}\,\text{eV}$ to $10^{-32}\,\text{eV}$.

\subsection{Modelling the Dilaton Field}

The classical dilaton field can be expanded as:
\begin{equation}
    \phi(\mathbf{x},t)=\sum_k\phi_{0,k} \cos(\omega_\phi t+\mathbf{k}\cdot\mathbf{x} + \varphi_k) \label{eq: dilaton_field}
\end{equation}
where $\omega_\phi=m_\phi c^2/\hbar$ is the Compton frequency and the $(\mathbf{k},\phi_{0,k},\varphi_k)$ are the momentum modes, associated amplitudes and phase of the field, which are determined by the dilaton DM power spectrum and local distribution described below. The phases $\varphi_k$ are fixed in a given physical realisation of this power spectrum.

The dilaton power spectrum, $P_\phi(k)$, and total matter power spectrum, $P(k)$, can be computed in linear cosmological perturbation theory using \textsc{axionCAMB}~\cite{axionCAMB}, a modified version of \textsc{CAMB}~\cite{Lewis:1999bs}. \textsc{axionCAMB} follows the procedure outlined above to follow the evolution of the field $\phi$ from adiabatic initial conditions, resulting in a prediction for $P(k,z)$~\footnote{\textsc{axionCAMB} is described as a model of axion DM, however the only assumption is that the scalar potential is $V(\phi)=m_\phi^2\phi^2/2$, which applies equally to pseudoscalars such as the axion, and scalars such as the dilaton, if self-interactions and other interactions are too weak to affect $P(k)$.} The resulting $P(k)$ can be used to generate a realization of the dilaton and CDM overdensity fields. This realization does not include the time evolution on the Compton scale, $\delta t\leq m_\phi^{-1}$. 

The sum over angular modes going from $\mathbf{k}$ to $k$ in $P(k)$ leads to a coherence time given by:
\begin{equation}
    t_c \approx \frac{\lambda_\mathrm{dB}}{v_\phi}=\frac{2\pi}{m_\phi v_\phi^2}
    \label{eqn:coherence_v}
\end{equation}
where the velocity is, in the case of a DM halo, approximated by the virial velocity.

The field $\phi$ evolves over three distinct timescales. On the longest time scales, the amplitude evolves over the scale of linear growth of structure in the Universe, i.e. over a Hubble time. This evolution is captured totally by the non-relativistic cosmological structure given by $P(k,z)$. 

Over the coherence time, Eq.~\eqref{eqn:coherence_v}, the amplitude also oscillates. In linear perturbation theory, oscillations over the coherence time are captured by the temporal oscillations in the linear growth factor and dilaton power spectrum (see e.g. Refs.~\cite{Marsh:2015xka,Hlozek:2014lca}). Oscillations in the linear power spectrum, while in principle captured by \textsc{axionCAMB}, are in practice ignored since we do not generate realizations of the density field using $P(k,t)$ sampled on such short time scales. As such, we take a stochastic approach to these scales, taking $\phi_{0,k}$ to be Rayleigh distributed, which can be derived analytically for virialized DM halos~\cite{Foster:2017hbq,Centers:2019dyn,Hui:2020hbq}. 

Finally, over the Compton time, $m_\phi^{-1}$, the amplitude oscillates. This oscillation is completely factored out in the non-relativistic approximation to structure formation. Noting that $H_0\approx 10^{-33}\text{ eV}$, we see that the Compton time, the linear growth time, and the Hubble time are all approximately equal for $m_\phi=10^{-33}\text{ eV}$, which occurs in quintessence models of dark energy, in which the entire Universe consists of a single coherent field. We do not treat this limit, since gauge issues arise when considering density perturbations on ultra large scales.

\begin{table}[]
    \centering
    \begin{tabular}{c|c|c}
        Mass & $\Omega_\phi / \Omega_{d}$ & Implementation\\ \hline
        $10^{-20}\,\mathrm{eV}$ & $1$ & Eq. \eqref{eq: voigt_high_mass} \\
        $10^{-22}\,\mathrm{eV}$ & $0.2$ & Eq. \eqref{eq: voigt_high_mass} \\
        $10^{-24}\,\mathrm{eV}$ & $0.2$ & Eq. \eqref{eq: voigt_medium_mass} \\
        $10^{-26}\,\mathrm{eV}$ & $0.03$ & Eq. \eqref{eq: voigt_medium_mass} \\
        $10^{-27}\,\mathrm{eV}$ & $0.03$ & Eq. \eqref{eq: voigt_medium_mass} \\
        $10^{-28}\,\mathrm{eV}$ & $0.02$ & Eq. \eqref{eq: voigt_small_mass} \\
        $10^{-29}\,\mathrm{eV}$ & $0.02$ & Eq. \eqref{eq: voigt_small_mass} \\
        $10^{-30}\,\mathrm{eV}$ & $0.02$ & Eq. \eqref{eq: voigt_small_mass} \\
        $10^{-31}\,\mathrm{eV}$ & $0.02$ & Eq. \eqref{eq: voigt_small_mass} \\
        $10^{-32}\,\mathrm{eV}$ & $0.06$ & Eq. \eqref{eq: voigt_small_mass} \\
    \end{tabular}
    \caption{Summary of the synthetic data sets that we test within this work. We present the mass, the dark matter fractions (taken from upper 1$\sigma$ limits presented in \cite{Lague2021,Kobayashi:2017jcf})} and which equations we use for computing the spectra.
    \label{tab: implementations}
\end{table}

\subsection{\label{ssec: ly-alpha-forest} Ly$\alpha$ forest}
The Ly$\alpha$ forest is an absorption feature occurring in the spectra of distant galaxies bluewards of the Ly$\alpha$ emission line as a sequence of densely packed, narrow absorption lines. These absorption lines are caused by the absorption of the illuminating light of the background quasar by the IGM. The observed flux, $F_\mathrm{obs}$, in the Ly$\alpha$ forest is often expressed as a normalized flux $F$:
\begin{align}
    F = \frac{F_\mathrm{obs}}{F_\mathrm{trans}},
\end{align}
where $F_\mathrm{trans}$ is the maximal flux that would have been observed at full transmission. The optical depth $\tau$ is defined by the logarithm of the normalized flux:
\begin{align}
    \tau = - \ln(F).
\end{align}

The optical depth is related to the neutral hydrogen density $n_\mathrm{HI}$ by convolution with the line emission profile, a thermal broadened Voigt profile $\mathcal{V}$ \cite{Bahcall1965, Gunn1965, Gallerani2006}:
\begin{align} \nonumber
    \tau(z_0) &= \sigma_0 c \int_\mathrm{LOS} dx(z) \frac{n_\mathrm{HI}(x, z)}{1+z} \\
    &\times \mathcal{V}\left( v_\mathrm{H}(z_0) - v_\mathrm{H}(z) - v_\mathrm{pec}(x, z), b_T(x, z), \gamma \right). \label{eq: optical_depth}
\end{align}
Here $\sigma_0$ is the effective Ly$\alpha$ cross section, $c$ the speed of light, $z$ and $z_0$ are denoting redshifts, $x(z)$ is the comoving distance at redshift $z$, $v_\mathrm{H}$ the differential Hubble velocity, $\gamma=\frac{\lambda_0}{2\pi\tau_{Ly\alpha}}$ (where $\lambda_0$ is the fiducial wavelength of the transition and $\tau_{Ly\alpha}$ the average time of transition) and $b_T$ the thermal broadening of the line. The thermal broadening parameter is proportional to the square root of the temperature $T$ of the IGM. In fact, it is \cite{Hui1997}:
\begin{align}
    b_T(x, z) = \sqrt{ \frac{2 k_\mathrm{B} T(x, z)}{m_\mathrm{p}} }, \label{eq: thermal_broadening}
\end{align}
where $k_\mathrm{B}$ is the Boltzmann constant and $m_\mathrm{p}$ the mass of the proton. 

In fact, the local IGM temperature depends on the overdensity again, see also our detailed discussion of IGM physics in Appendix \ref{sec: igm-physics}. Hence, Eq. \eqref{eq: optical_depth} cannot be understood as a true convolution as the emission profile depends on the neutral hydrogen density again due to the thermal broadening of the line. We will use the term convolution nevertheless for the remainder of the paper.

\subsection{\label{ssec: mod-voigt-profile} Dilaton Modified Voigt profile}

\begin{figure*}
    \centering
    \includegraphics[width=\textwidth]{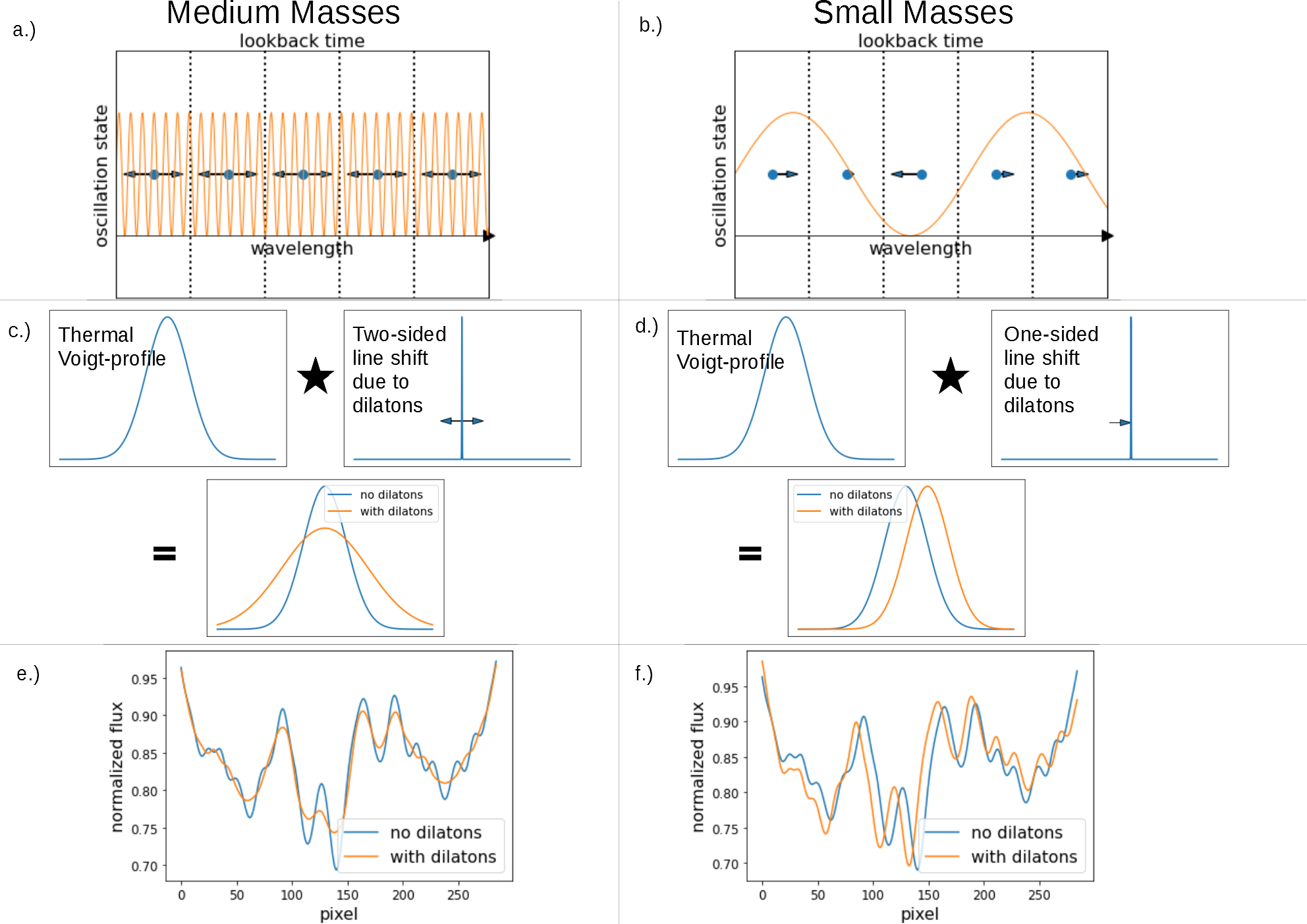}
    \caption{Summary of the effect of dilaton DM on the Ly$\alpha$ forest. We show the broadening effect that appears for medium masses in the left panels, and the shifting effect that is dominant at the smallest masses in the right panels. \textit{Subfigure A}: We show the pixels in the observation/simulation as a function of apparent absorption wavelength/redshift (i.e. of the distance along the line of the sight to the quasar) with dashed vertical line. The blue circles represent hydrogen atoms spread across the pixel. Moreover, we show in orange the oscillation state $\phi_m$ of the dilaton (and thus $\alpha$) as a function of lookback time, i.e. of the travel time of a photon that travels along a line of sight. For medium masses, the dilaton goes through several oscillations during pixel crossing, causing additional blueshift and redshift. The absorption profile caused by a delta function neutral hydrogen density profile sitting in one of the pixels is broadened to smaller and higher apparent wavelength (arrows). \textit{Subfigure B}: The same as Subfigure A, but for the smallest masses. The oscillation period of the dilaton is smaller than the length of the pixel. Some of the pixels are blueshifted and some are redshifted (indicated by the arrows), depending on the current status of the dilaton oscillation. \textit{Subfigure C}: The resulting absorption profile of a delta function point source as a convolution of the thermal Voigt profile and an additional dilaton (oscillating) shifting of the profile (arrows) resulting in additional broadening.  \textit{Subfigure D}: For low masses, dilaton DM leads to a shift of the Voigt profile. \textit{Subfigure E}: A small fraction of the resulting spectrum with and without the dilaton (broadening) effect. \textit{Subfigure F}: Same as Subfigure E, now with a visible displacement of the absorption lines (note that effects in E and F are greatly exaggerated).}
    \label{fig: sketch}
\end{figure*}

We present now in this subsection how the dilaton affects the absorption features in the Ly$\alpha$ forest, i.e. how it affects the Voigt profile. We present a comprehensive illustration of the effects that we are looking for in mock data in Fig.~\ref{fig: sketch}.

The absorption feature (in frequency space) of a neutral (ground state) hydrogen gas (HI) at a given temperature and redshift is described by a Voigt profile (convolution of Gaussian and Lorentzian profiles) centred around the $n=2\rightarrow n=1$ transition wavelength, adjusted for redshift \cite{Mitchell.textbook}:
\begin{align} \nonumber
    \mathcal{V}(v_H(x,z),&b_T(x,z), \gamma) = \\
    & \frac{\gamma}{\pi^{3/2}b_T} \int_{-\infty}^\infty \frac{e^{-\frac{v'^2}{b_T}}}{\gamma^2 + (v_H(x,z)-v')^2}dv'
\end{align}
where the numerator comes from a thermal distribution of velocities and the denominator is a Lorentzian \cite{Hoyt:1930} which comes from the finite time of the transition, i.e. $\gamma=\frac{\lambda_0}{2\pi\tau_{\text{Ly}\alpha}}$ (where $\lambda_0$ is the fiducial wavelength of the transition, $\tau_{\text{Ly}\alpha}$ is the characteristic time of the Ly$\alpha$ transition).

We remind ourselves of the energy of the $n=2\rightarrow n=1$ transition (in natural units):
\begin{equation}\label{eq: Lyalpha energy}
    \Delta E_{\text{Ly}\alpha} = \frac{3}{4} \text{Rydberg} = \frac{3m_e\alpha^2}{8} 
\end{equation}

The transition energy now locally changes by a small amount $\delta \Delta E = 2 \Delta E \delta \alpha / \alpha \propto d_i\phi(\mathbf{x},t)$ where $d_i=d_e,d_{m_e}$, see Eq. \eqref{eq: fine-structure}.\footnote{For Ly$\alpha$ forest searches we neglect further corrections from local spin temperatures. However, in principle these can be included here, once more precise 21 cm data is available for modelling, similarly to how temperature is incorporated.} The modified Voigt profile (to first order) is:
\begin{align} \nonumber
    \mathcal{V}(v_H(x & ,z),b_T(x,z),\gamma) = \frac{\gamma}{\pi^{3/2}b_T}\int_{-\infty}^\infty \\ &
    \frac{e^{-\frac{v'^2}{b_T}}}{(\gamma^2 + (v_H(x,z)-v' - 2c\kappa d_i \phi_r\phi_m)^2)} dv',
    \label{eq: voigt_small_mass}
\end{align}
where $\phi_m(x, z)$ is a dimensionless variable that describes the current state of the sinusoidal oscillation and $\phi_r(x, z)$ (dimension of energy) is the amplitude of the dilaton field, which is subject to stochastic fluctuations around the mean value $\langle\phi_r\rangle=\sqrt{2 \rho_\phi(\mathbf{x},t)}/m_\phi$ \cite{Foster:2017hbq}.

We consider the absorption profile to be built up of pixels. Each pixel is traversed by the quasar light in a time $t_{\rm pix}$. If the beam of photons which will eventually reach our telescope and travelling through the distance equivalent to one pixel (i.e. $\sim23$~kpc for UVES SQUAD \cite{Murphy2019} spectral resolution) saw at least one oscillation of the dilaton field during that time, we have to average the oscillation state, compare also the illustration in panel a.) of Fig.~\ref{fig: sketch}:
\begin{align} \nonumber
    & \mathcal{V}(v_H(x,z),b_T(x,z),\gamma) = \frac{\gamma}{\pi^{5/2}b_T}\int_{-1}^1\int_0^\infty \\ &
    \frac{e^{-\frac{v'^2}{b_T}}}{(\gamma^2 + (v_H(x,z)-v' - 2c\kappa d_i \phi_r\phi_m)^2)\sqrt{1-\phi_m^2}}d\phi_m dv', 
    \label{eq: voigt_medium_mass}
\end{align}
where the $\frac{1}{\sqrt{1-\phi_m^2}}$-factor appears as the derivative of the arcsin-function.

The local coherence time is estimated by Eq. \eqref{eqn:coherence_v}. We estimate the average velocity as the Zel'dovich velocity \cite{Zeldovich1970}, giving  for moderate masses $10^{-20}\,\mathrm{eV} \lesssim m_\phi \lesssim 10^{-26}\,\mathrm{eV}$ velocities $v_\phi \approx 1000\,\mathrm{km/s}$, i.e. $\langle v_\phi \rangle^2/c^2 \approx 10^{-5}$. For smaller masses the velocity drops, e.g. to the order of $1\,\mathrm{km/s}$ for mass $m_\phi = 10^{-30}\,\mathrm{eV}$. If $t_{\rm pix}>t_{\rm c}>t_m$, then the Compton scale oscillations and coherence scale fluctuations can both be averaged over. The modified Voigt profile is then given by:
\begin{align} \nonumber
    & \mathcal{V}(v_H(x,z),b_T(x,z),\gamma) = \frac{\gamma}{\pi^{5/2}b_T}\int_{-1}^1\int_0^\infty\int_{-\infty}^\infty \\ &
    \frac{e^{-\frac{v'^2}{b_T}}\left(\frac{2\phi_r}{\langle \phi_r \rangle}e^{-\frac{\phi_r^2}{\langle \phi_r \rangle^2}}\right)}{(\gamma^2 + (v_H(x,z)-v' - 2c \kappa d_i \phi_r\phi_m)^2)\sqrt{1-\phi_m^2}}d\phi_m d\phi_rdv' \label{eq: voigt_high_mass}
\end{align}
where $\phi_r$ is a Rayleigh distributed variable \cite{Foster:2017hbq}, with mean fixed by the expectation value of the dilaton density over many coherence times, i.e. $\langle\phi_r\rangle=\sqrt{2 \rho_\phi(\mathbf{x},t)}/m_\phi$.

That leaves us with three different mass regimes:
\begin{enumerate}
    \item In the extreme case where $t_{\rm pix}<t_m<t_c$ \footnote{This is approximately $10^{-32}$~eV, below which field oscillation freezes and are no longer dynamical. However spatial variations of non oscillating could still be detected, similarly to \cite{VanTilburg:2015oza}.} we must deterministically evolve the field on the Compton time with a fixed global phase. This leads to shifts in the central value of the Voigt profile from pixel to pixel, oscillating from blueshifted to redshifted on cosmological timescales (see right panels in Fig.~\ref{fig: sketch}). Hence, we compute the Voigt profile by Eq.~\eqref{eq: voigt_small_mass} explicitly as long as $t_\mathrm{pix} << t_m$.
    \item At intermediate masses (i.e. if $t_m<t_{pix}<t_c$), this gradually gives way to many oscillations inside a single pixel. We have to average the oscillation term and compute the Voigt profile by Eq. \eqref{eq: voigt_medium_mass}. In this case the Voigt profile is simply broadened, but the broadening is not fixed deterministically by $\rho_\phi$.~\footnote{In principle this step could be computed deterministically by recomputing $P(k,t)$ on the pixel crossing time, but this would be computationally prohibitive.} In this case we take a draw for $\phi_r$ from the Rayleigh distribution for each coherence time travelled along the line of sight, and rescale it according to the local dilaton density. This scheme is approximate: a complete picture could be found only by full simulation of the field on sub-coherence timescales (e.g. Refs.~\cite{schive2014cosmic,Mocz:2019pyf}, although approximate methods such as Ref.~\cite{Lague:2020htq} might suffice on the quasi-linear scales probed). Our treatment of this case is inspired by Ref.~\cite{Centers:2019dyn} who consider the same regime of scales for direct detection experiments.
    \item  As we increase the dilaton mass further (i.e. if $t_m<t_c<t_{pix}$), the k-modes of the field's overdensities start to decohere in the pixel crossing time, at which point we need to average over a Rayleigh distribution with average being the classical field amplitude. Moreover, we still have to average the oscillation term. Hence, we compute the Voigt profile by Eq. \eqref{eq: voigt_high_mass}. The dilaton oscillations broaden the Voigt profile more so than the previous case, as shown in Subfigure c.) of Fig.~\ref{fig: sketch}.
\end{enumerate}

\section{Measuring the Dilaton Coupling}

We have demonstrated in the previous subsection that dilatons affect the Ly-$\alpha$ forest by line broadening and line shifting depending on the dilaton mass. Regarding these phenomena, one may ask how detection of these is affected by possible degeneracies with similar effects or underlying uncertainties in our assumptions (e.g. the underlying HI density field is unknown, temperature modelling and redshift space distortions introduce degeneracies and the dilaton dark matter fraction is unknown as well). Detailed study of the cross-correlations between various sources of inputs to our fitting of Ly-$\alpha$ survey data is beyond the scope of this study, nevertheless we discuss here some of the main sources of degeneracies, the uncertainties it generates and several multiwavelength avenues we can use to compensate and strengthen the statistical significance of the dilaton effect. We sketch our proposed analysis strategy on observational data in Fig.~\ref{fig: strategy}.

\subsection{Dilaton dark matter fraction}\label{ssec: multimessenger}
In this work we are considering placing limits to dilaton couplings $d_i$ \emph{assuming} input values of $(m_\phi,\Omega_\phi)$. How would one measure $d_i$, or set a limit to it, in practice? We have not considered simultaneous variation of $(d_i,\Omega_\phi)$, but it is likely that the parameters are highly degenerate if both were reconstructed simultaneously using Ly$\alpha$. Thus, \emph{measuring} $d_i$ requires use of multiple probes to break this degeneracy and establish a concordance cosmology. 

At present, the CMB anisotropies and other cosmological probes set upper limits to $\Omega_\phi$ at fixed $m_\phi$, as reported in Table~\ref{tab: implementations}. If a simultaneous fit were performed to Ly$\alpha$ and existing CMB data varying both $\Omega_\phi$ and $d_i$, the degeneracies would not be broken, and in a combined Bayesian analysis one would find an upper limit to $d_i$ marginalised on $\Omega_\phi$, and correspondingly weaker than the limits shown in Figs.~\ref{fig:de_constraints} \& \ref{fig:dm_constraints}.~\footnote{Frequentist analysis of CMB data is extremely uncommon. However, we mention that a profile likelihood on $\Omega_\phi$ peaked at zero would lead to no limit to $d_i$ from Ly$\alpha$.}

Future CMB measurements of both primary~\cite{Hlozek:2016lzm} and secondary~\cite{Farren:2021jcd} anisotropies, intensity mapping~\cite{Bauer:2020zsj}, and galaxy surveys~\cite{Marsh:2011bf}, can improve these limits by several orders of magnitude, allowing for possible detection of non-zero $\Omega_\phi$ at high significance for DM fractions around 1\% percent. These probes are, compared to the Voigt profile broadening considered here, insensitive to $d_i$, and a detection of $\Omega_\phi\neq 0$ would leave unanswered the question what gravitational forces $\phi$ possesses.

However a detection of non-zero $\Omega_\phi$ in one of the above mentioned probes would break the degeneracy between $d_i$ and $\Omega_\phi$ in Ly$\alpha$, and also give a theoretical prediction for the matter power spectrum, $P(k)$. The power spectrum can furthermore be reconstructed using our inversion procedure \cite[see][]{Mueller2020} for the Ly$\alpha$ forest, which would allow for the establishment of concordance, and measurement of $d_i$, as outlined in the next subsection. The present analysis for dilaton-like couplings is thus very similar in spirit to the analysis of \citet{Fujita:2020aqt} for measuring axion-like couplings via birefringence.

\subsection{Approximate estimation strategy}
In real observations the underlying density field is unknown. This poses a particular problem for the estimation procedure since we need to have a criterion to unveil that the absorption line was broadened/shifted by coupling to dilatons instead of the underlying overdensity in the IGM just having a larger width (to explain broadening) or being placed at another redshift distance (to explain line shifting). In concordance with the procedure described in detail in \citet{Mueller2021} we propose that this degeneracy could be lifted by using the matter power spectrum as a consistency check.

As demonstrated in \citet{Mueller2020} the matter distribution along a single line of sight can be recovered best with the iterative Gauss-Newton method (IRGNM) proposed by \citet{Pichon2001}. Starting from an initial guess, the residual fit to the observed data is minimized by forward modelling in a Newton-type optimization scheme. An alternative approach, while less precise, would be the PC \citep{Gallerani2011} and RPC \citep{Mueller2020} methods that are independent of the thermal model. However, we are assuming the thermal model to be known in this work and thus stick to the more precise IRGNM procedure.

\begin{figure*}
    \centering
    \includegraphics[width=\textwidth]{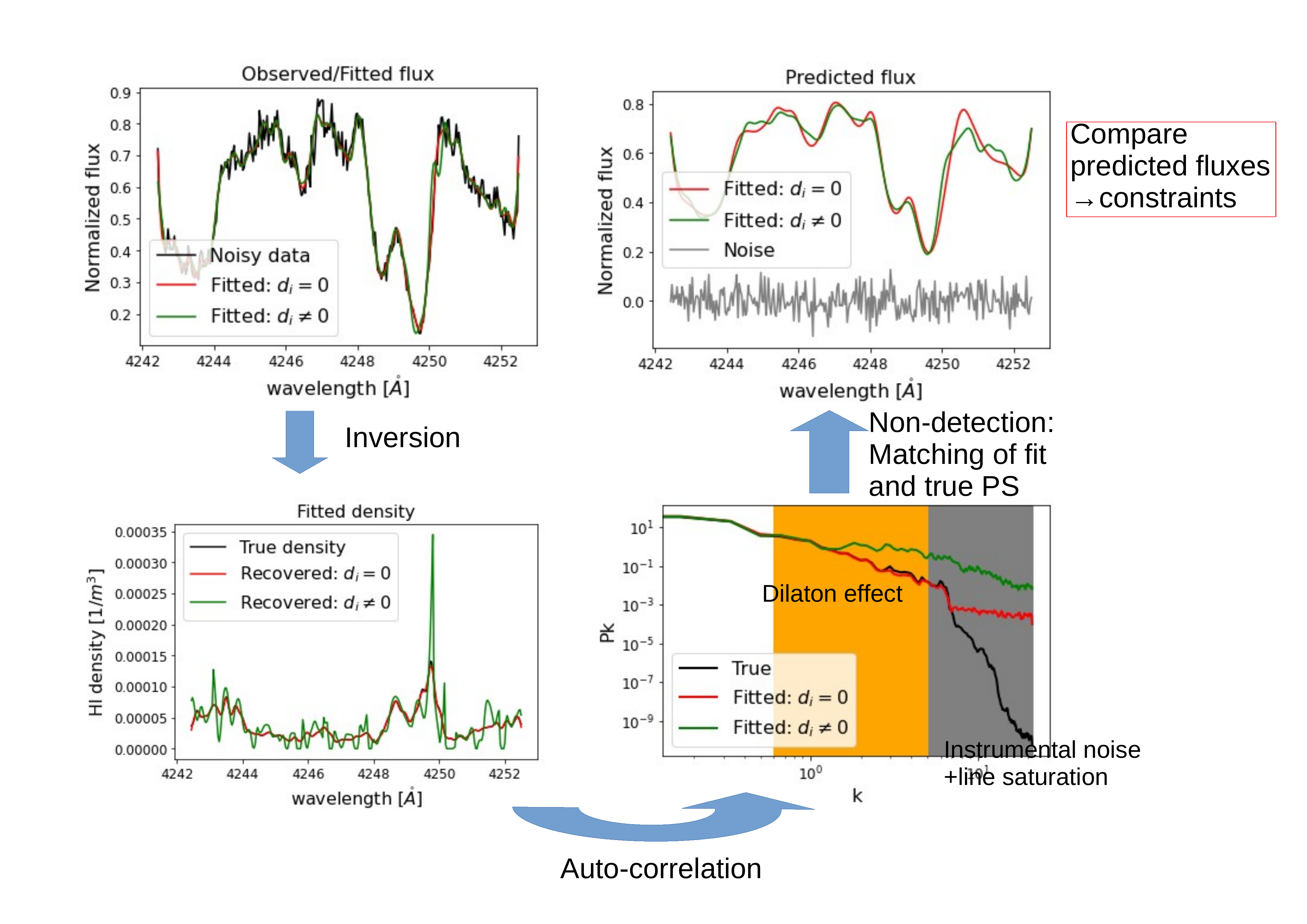}
    \caption{Observational strategy for studying real data. Note that the true $m_\phi$, $\Omega_\phi$ and thus $P(k)$ are assumed known e.g. from an independent cosmological probe. Example shown for $m_\phi = 10^{-24}\,\mathrm{eV}$ and $d_i = 10^{4}$ in the case of non-detection (i.e. observed spectrum with fixed dilaton mass and without dilaton coupling): for a fixed mass we invert the observed spectrum with a dilaton coupling $d_i = 0$ (red line) and $d_i \neq 0$ (green line). Both fits are successful in modelling the flux (upper left panel, red and green line match and describe the expected noise-free flux), but the recovered HI density is different. The fit with $d_i = 0$ matches the true density profile, for the fit with $d_i \neq 0$ we overfit the variation in the density profile which gives a spurious signal on the high wavenumber modes of the matter power spectrum of the recovered density profiles (lower right panel). In the high wavenumber regime we can identify the (falsely assumed) broadening effect of dilatons by a discrepancy between observed and simulated power spectrum (lower right panel, orange shaded region). If we would have observed a mismatch between theoretical power spectrum and the $d_i = 0$ fit, we would have a detection of a non-zero coupling, see Fig.~\ref{fig: strategy2}. In the other case (non-detection, i.e. the fit without dilaton coupling matches the theoretical power spectrum) presented here, the recovered density approximates the correct power spectrum sufficiently well (see lower left panel). Hence, we assume that the recovered density profile describes the correct one and we compute constraints by predicting the observed flux from the recovered density (fitted with $d_i = 0$) for various dilaton couplings $d_e \neq 0$ and compare the fluxes to the $d_i = 0$ prediction (upper right panel). The difference between the predicted fluxes has to be compared to the instrumental noise of the observation (grey).}
    \label{fig: strategy}
\end{figure*}

\begin{figure*}
    \centering
    \includegraphics[width=\textwidth]{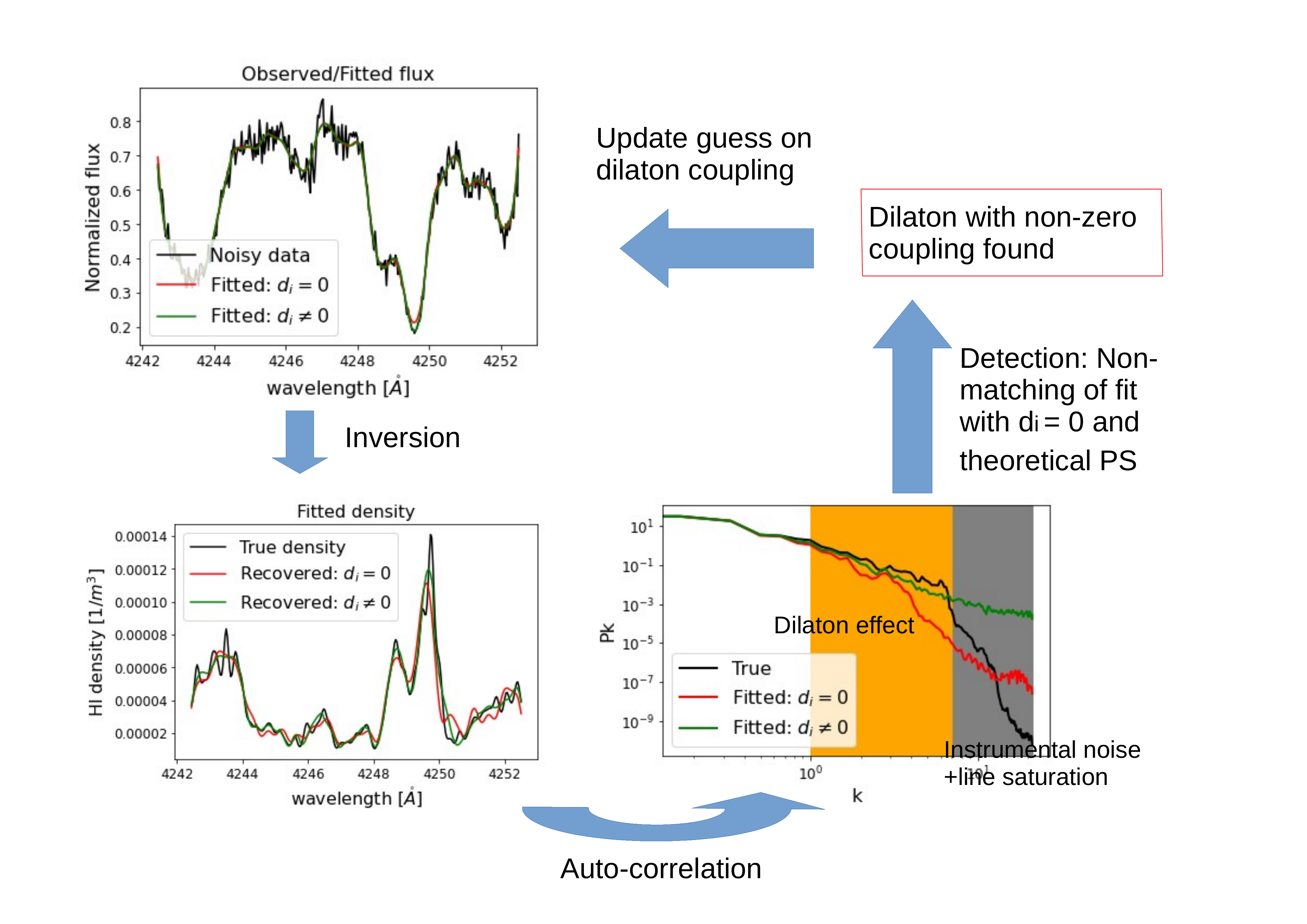}
    \caption{The same as Fig.~\ref{fig: strategy}, but for the case of detection. We invert the observed spectrum with a guess dilaton coupling starting with $d_i = 0$. We compute the power spectrum of the recovered density profile by auto-correlation of the recovered density profile and compare to the theoretical prediction. If the fit and the true power spectrum match, we would have detection and would proceed as sketched in Fig.~\ref{fig: strategy}. In the case of non-matching we would have a detection of a non-zero coupling, i.e. the observed spectrum is not consistent with $d_i = 0$. We iteratively take a next greater guess for $d_i$, compute the inversion and compare to the theoretical power spectrum until consistency is achieved.}
    \label{fig: strategy2}
\end{figure*}

We invert the observed spectrum (i.e. recover the density field) assuming various dilaton couplings $d_i$ (at fixed $m$ and fixed $\Omega_\phi$) and inspect recovered densities for their power spectrum. This strategy is outlined in Fig.~\ref{fig: strategy} in the case of non-detection and in Fig.~\ref{fig: strategy2} in the case of detection. In the case of detection, such that the dilaton effect is larger than the noise in the Ly$\alpha$ flux, then only the reconstructed $P(k)$ from Ly$\alpha$ with this $d_i$ (within some error) will agree with the theoretical prediction for $P(k)$ from the assumed non-zero $\Omega_\phi$, see lower panels in Fig.~\ref{fig: strategy2}. 
Conversely, in the case of non-detection, such that the dilaton effect is smaller than the noise in the Ly$\alpha$ flux, the agreement between predicted and reconstructed $P(k)$ would allow an upper limit to $d_i$ to be set as we are recovering the correct HI density profile from the Ly-alpha flux (up to errors induced by the flux noise)
while setting $d_i=0$. If a dilaton broadening/shifting effect is (falsely) imposed during reconstruction, the recovered density would overfit the small scale variation in the density (lower left panel in Fig.~\ref{fig: strategy}) affecting the power spectrum (lower right panel in Fig.~\ref{fig: strategy}.

A full statistical analysis would have to perform a joint estimation with these two observables. However, in the rest of the study, we will focus on a scenario of absence of dilaton signal (exclusion). For this particular situation of non-detection we developed an approximate, shortened estimation procedure outlined in Fig.~\ref{fig: strategy} that is sufficient for the scope of this work.

Assuming that the dominant source of noise is instrumental and that the dilaton effect on the flux is smaller than the instrumental noise, the true HI density can be uniquely identified (within the uncertainty induced by the instrumental noise) by constraining the reconstruction to reproduce a $\Lambda$CDM power spectrum (in the absence of dilatons, reconstructions with falsely imposed dilatons on the other hand would fail the $\Lambda$CDM test). Once the true HI density is estimated with high fidelity, we can place constraints on the dilaton coupling by looking for the largest dilaton effect still compatible with the data, i.e. by predicting the dilaton effect on the Ly$\alpha$ forest from the known (recovered) HI overdensity and comparing the predicted flux in the non-coupled and the coupled cases (upper right panel in Fig.~\ref{fig: strategy}. This also resembles the method that was successfully applied by \cite{Mueller2021} for measuring the mean temperature of the IGM.

This discussion however potentially reveals an additional source of systematic noise when determining the coupled and uncoupled profiles: uncertainty in the matter power spectrum. Throughout this study we implicitly assume the Ly$\alpha$ survey noise to be the dominant nuisance parameter in our likelihoods, but we may add other sources, such as the uncertainties in the $\Lambda$CDM parameters determined by \emph{Planck}.


\subsection{Astrophysical degeneracies}
The absorption profile of Ly$\alpha$ forest lines is not only affected by the profile of the underlying density profile and the dilaton coupling, but also by thermal broadening, see Sec. \ref{ssec: ly-alpha-forest}. The effect of temperature is complex: the absorption lines are not only Doppler broadened, higher temperatures also smooth the gas. Moreover, the temperature is not constant. In particular unlike with dilaton it is correlated with density, creating deeper absorption features at higher temperatures. We refer the reader to our discussion in Appendix \ref{sec: igm-physics} for details of our temperature model.

Measuring and modelling the temperature of the IGM at redshift $z\sim 2-3$ is a currently very active field of research \cite{Becker2011, Garzilli2012, Boera2014, Hiss2018, Telikova2019, Walther2019, Gaikwad2020, Mueller2021}. While these methods seem to coincide at temperatures between $10000\,\mathrm{K}$ and $15000\mathrm{K}$ with errors of only several hundreds of $K$ (i.e. SNR$>20$), which would justify ignoring them, there is still some discrepancy between the various observations. We expect consistency of future studies to improve, however it is possible the uncertainties in the IGM temperature could limit the predictive power of Ly$\alpha$ forest surveys.
Nevertheless thermal broadening and dilaton broadening are not the same (Eq. \ref{eq: voigt_high_mass}). We stress the study of possible degeneracies and correlations warrants more detailed study which we leave to future work. In the rest of this work we select a thermal model (see Appendix \ref{sec: igm-physics}) and ignore thermal uncertainties.
Most importantly, we note that limits on dilaton couplings only exceed laboratory bounds in the region of line shifting, where the absorption lines are displaced instead of broadened and there is no degeneracy with thermal effects.


For the rest of this study we assume that all astrophysical uncertainties are accounted for and we are ignoring them in order to find forecasts.

\subsection{Multi-probe forecasts}
It is beyond the scope of the present work to make a multi-probe forecast for simultaneous measurement of $(d_i,\Omega_\phi)$, but we note that to compensate these uncertainties we may combine the present analysis with the Fisher matrix of Ref.~\cite{Bauer:2020zsj} or other probes. Multi-probe forecasts can improve statistical significance of dilaton bounds if used to cross check relative peak shifts from absorptions with different dependence on $\alpha$, as is the case of the Ly$\alpha$ and 21 cm lines (see Sec. \ref{sec: results}). In the absence of correlations, 21 cm lines would act as an independent measurement of the location of the peak. Further, cosmic standard candles in the field (e.g. in a protocluster) with luminosity distances of galaxies along the line of sight (LOS) of Ly$\alpha$ studies may be used to check the redshift of emission peaks as well. Ly$\beta$ forest lines could also provide a measurement: although the peaks of the absorption lines are displaced by the same factor as Ly$\alpha$, we look for variation in the ratio of the Ly$\alpha$ and Ly$\beta$ absorption peak wavelengths occurring from the same overdensity.

Finally, it is worth noting briefly that for the lightest masses ($m_\phi\lesssim 10^{-28}$~eV), in the case of misalignement at CMB decoupling, a CMB spectral distortion would be induced, with possible limits to $d_i$ using the COBE~\cite{Smoot:1998jt} spectrum. On the other hand, if $\phi$ were axion-like, CMB polarization could provide simultaneous measurement of $\Omega_\phi$ and the axion-photon coupling via birefringence~\cite{Hlozek:2016lzm,Fujita:2020aqt}.

\section{\label{sec: software}Software/Synthetic Data}

\subsection{\label{ssec: reglyman} Reglyman}

Based on the power spectra computed with \textsc{axionCAMB} we simulate the neutral hydrogen IGM overdensity by the lognormal approach. In the lognormal approach the IGM overdensities are modelled by a random lognormal distribution with an auto-correlation function specified by the power spectrum. In a nutshell, we start with a Gaussian white noise field and multiply it with the matter power spectrum. The Fourier transform of this field, also interpreted as the linear density perturbation, is a Gaussian distributed random field with the correct auto-correlation. We project the linear density perturbation to the quasi-linear regime by taking the exponential, i.e. the resulting overdensity field is a lognormal distributed random variable. This semi-analytic approach has a wide range of applications \cite{Coles1991, Bi1992, Bi1997, Choudhury2001, Viel2002, Gallerani2006, McDonald2006, Hand2018, Ribera2012, Mueller2020, Karacayli2020, Mueller2021} as it allows for fast computations of large boxes and is built in the publicly available NBODYKIT \footnote{Publicly available under https://nbodykit.readthedocs.io/en/latest.} software package \cite{Hand2018}. However, it is known that the lognormal model might be inadequate for overdensities at very small scales and highly non-linear overdensities \cite{Choudhury2005} although this is believed to be a smaller problem to the distribution of ordinary matter due to pressure smoothing \cite{Gallerani2006}. For a deeper discussion of the lognormal approach we refer to the discussion in our previous publication \cite{Mueller2021}.   

We implemented the lognormal approach in our software package \textsc{REGLYMAN} \cite{Mueller2020, reglyman} by interfacing to \textsc{NBODYKIT}. \textsc{REGLYMAN} is strongly based on the publicly available \textsc{REGPY} code for inverse problems in general \cite{regpy}. For this work for the forward simulation of the flux we ignore peculiar velocities that are accessible from the lognormal model from the Zel'dovich approximation \cite{Zeldovich1970, White2014}. We computed the baryonic power spectrum in previous works from the dark matter power spectrum by pressure smoothing with the Jeans scale as described in \cite{Fang1993, Choudhury2001, Gallerani2006, Zaroubi2006}. However, in the present work we take the baryon power spectrum directly from axionCAMB (since the baryons are also indirectly affected by the dilaton sound speed).

\begin{figure*}
    \centering
    \includegraphics[width=0.8\textwidth]{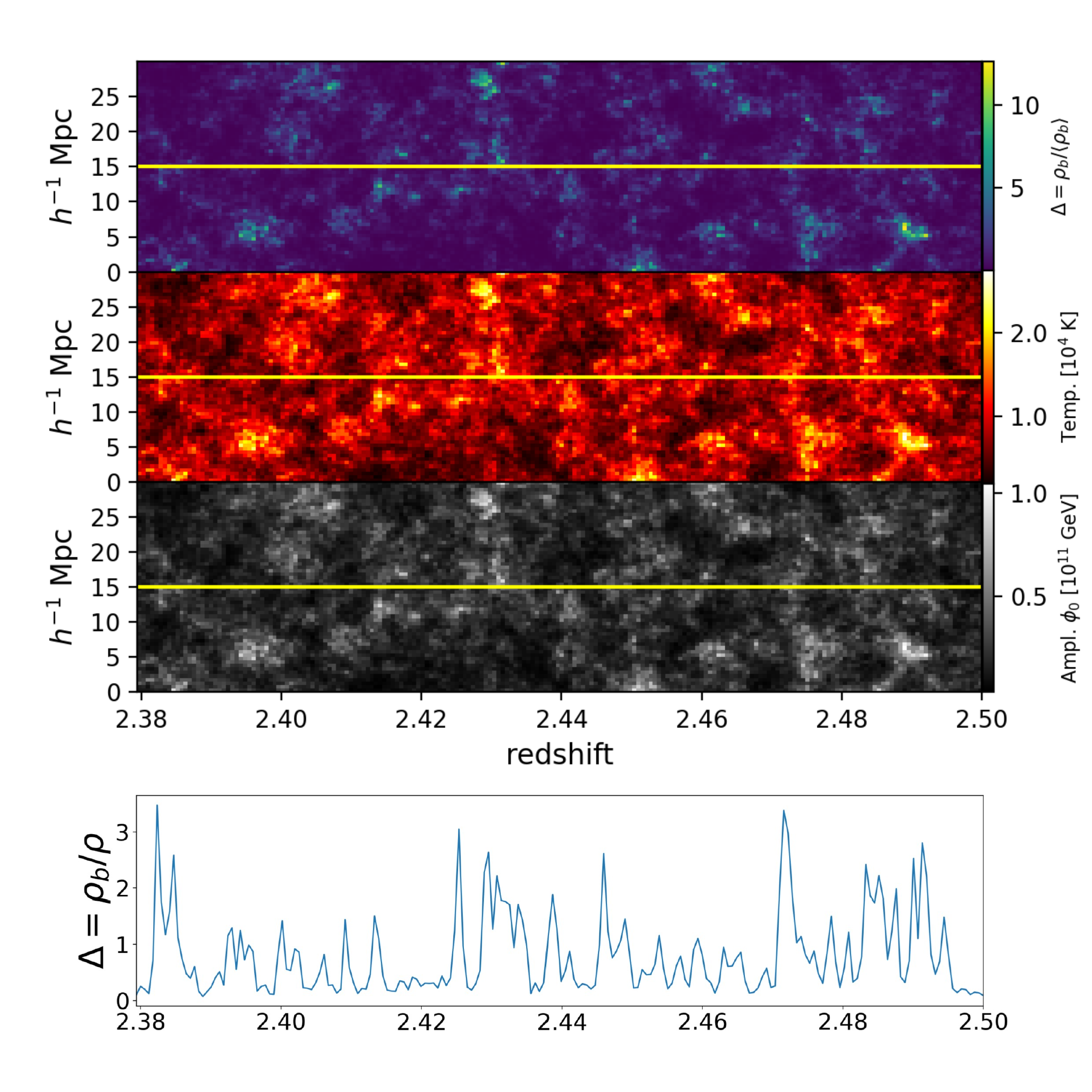}
    \caption{2D slice through a 3D box simulated with \textsc{REGLYMAN} and \textsc{axionCAMB}. The box shown in this figure is simulated with $m_\phi = 10^{-24}\,\mathrm{eV}$ and density in Table \ref{tab: implementations}. The upper panel shows the baryonic density perturbation, the upper middle panel the temperature in the IGM at the indicated redshift, and the middle bottom panel the amplitude of the dilaton field. We draw lines of sights (yellow) from the box. The baryon density profile along the line of sight is presented in the bottom panel.}
    \label{fig: simulation}
\end{figure*}

We present a two dimensional slice through a very small simulation box computed with \textsc{REGLYMAN} and \textsc{axionCAMB} with a dilaton mass of $m_\phi = 10^{-24}\,\mathrm{eV}$ in Fig.~\ref{fig: simulation}. However, for predicting the Ly$\alpha$ flux we used a larger box (more lines of sight that can be assumed to be uncorrelated) with a much greater resolution in the direction along the line of sight. The upper panel shows the computed baryonic density perturbation $\Delta$. The density perturbation consists of diffuse overdense regions and voids (underdensities). These structures are tracked in the Ly$\alpha$ forest. The middle panel shows the current IGM temperature (i.e. the amount of thermal broadening). The temperature is proportional to the the density perturbation, as discussed in Appendix \ref{sec: igm-physics}. We assume here that the temperature at mean density $T_0(z)$ is constant over the small redshift bin under consideration. We show in the bottom panel the dilaton amplitude which is closely correlated with the baryon density perturbation as well. For the purpose of computing the dilaton field amplitude from $\Delta$ we first project the baryon density to the dark matter overdensity by linear biasing, i.e. by forwarding every spatial mode separately by the relation of the dark matter and ordinary matter power spectra. Then we compute the dilaton density by the assumed dark matter fraction of dilatons and the amplitude by $\langle\phi_r\rangle=\sqrt{2 \rho_\phi}/m_\phi$.

With a fixed temperature at mean density $T_0$, a fixed mean density $\hat{n}_\mathrm{HI}$ and a fixed phototropic index $\zeta$, we use Eq.~\eqref{eq: optical_depth} to compute the optical depth in the Ly$\alpha$ forest, the evolution of the temperature is computed according to Appendix \ref{sec: igm-physics}. In the case of a non-vanishing coupling $d_e$ we replace the Voigt profile by the modified Voigt profile in Eq. \eqref{eq: voigt_small_mass}, \eqref{eq: voigt_medium_mass} or \eqref{eq: voigt_high_mass}.

\subsection{\label{ssec: synthetic-data} Synthetic data}

To produce the bounds in Figs.~\ref{fig:de_constraints} \& \ref{fig:dm_constraints} we create noisy mock Ly-$\alpha$ spectra (data) from exact hydrogen overdensities. We use the exact overdensities as our fit for $\mathcal{H}_1$ (see Sec.\ref{ssec: likelihood}) to the data resembling the last step of our full pipeline presented in Sec. \ref{ssec: multimessenger}. We assume these (exact) overdensities are noisy in Sec.\ref{ssec: likelihood} to mimic the typical noise from inversion.

We then simulated the dilaton effect in a wide range of masses (from $10^{-20}\,\mathrm{eV}$ down to $10^{-32}\,\mathrm{eV}$) and couplings. There are already existing bounds on the fraction of ultralight particles from the CMB \cite{Hlozek2015, Lague2021} and from galaxy clustering \cite{Lague2021}. We chose dark matter fractions that are in agreement with the combined $95\%$ exclusion limits of \cite{Lague2021,Kobayashi:2017jcf}. We present in Tab. \ref{tab: implementations} the masses assumed for our test runs and the corresponding fractions $\Omega_\phi / \Omega_{DM}$. 

As the broadening effect could be small, we modelled our synthetic data on high-resolution and high quality tomographic data. In fact, we mimicked the spectral resolution and signal-to-noise ratio (SNR) of the UVES SQUAD survey \cite{Murphy2019}. UVES SQUAD is a fully reduced spectroscopic survey of 467 QSO's starting at very low redshift and up to redshift 5. The quasars have a median continuum to noise ratio of $20$ at a spectral resolution of $2.5\,\mathrm{km/s}$ at a wavelength of $5500\,\mathring{A}$ \cite{Murphy2019}. We used a simplified set of synthetic data where we assumed that all lines of sight are at the same redshift $z=2.5$ and all spectra had the same spectral resolution $2.5\,\mathrm{km/s}$ and same continuum to noise level $20$. We applied the common noise model that was used previously for inversion problems in the Ly$\alpha$ forest \cite{Pichon2001, Mueller2020}, i.e.:
\begin{align}
    \sigma^2 = \frac{F^2}{SNR^2} + \sigma_0^2,
\end{align}
where $\sigma_0$ dominates the noise distribution for very small fluxes. We took $\sigma_0 = 0.005$. We created synthetic spectra with rest frame wavelengths between Ly$\alpha$ at $1216\,\mathring{A}$ and Ly$\beta$ at $1025\,\mathring{A}$. To utilize parallel computations we sliced every spectrum in slices of roughly $10\,\mathring{A}$ in length similarly to previous works \cite{Mueller2021}. 

At redshift $z=2.5$ the pixel size corresponds to $\sim 23\,h^{-1}\mathrm{kpc}$ in comoving length corresponding to a time scale $t_\mathrm{pix} \sim 3.5 \cdot 10^{12}\,\mathrm{s}=1.9\times 10^{-28}$~eV, which roughly separates the dilaton masses into line broadening and line shifting cases, see Sec.\ref{ssec: mod-voigt-profile}.

\begin{figure*}
    \centering
    \subfigure[]{
        \includegraphics[width=0.45\textwidth]{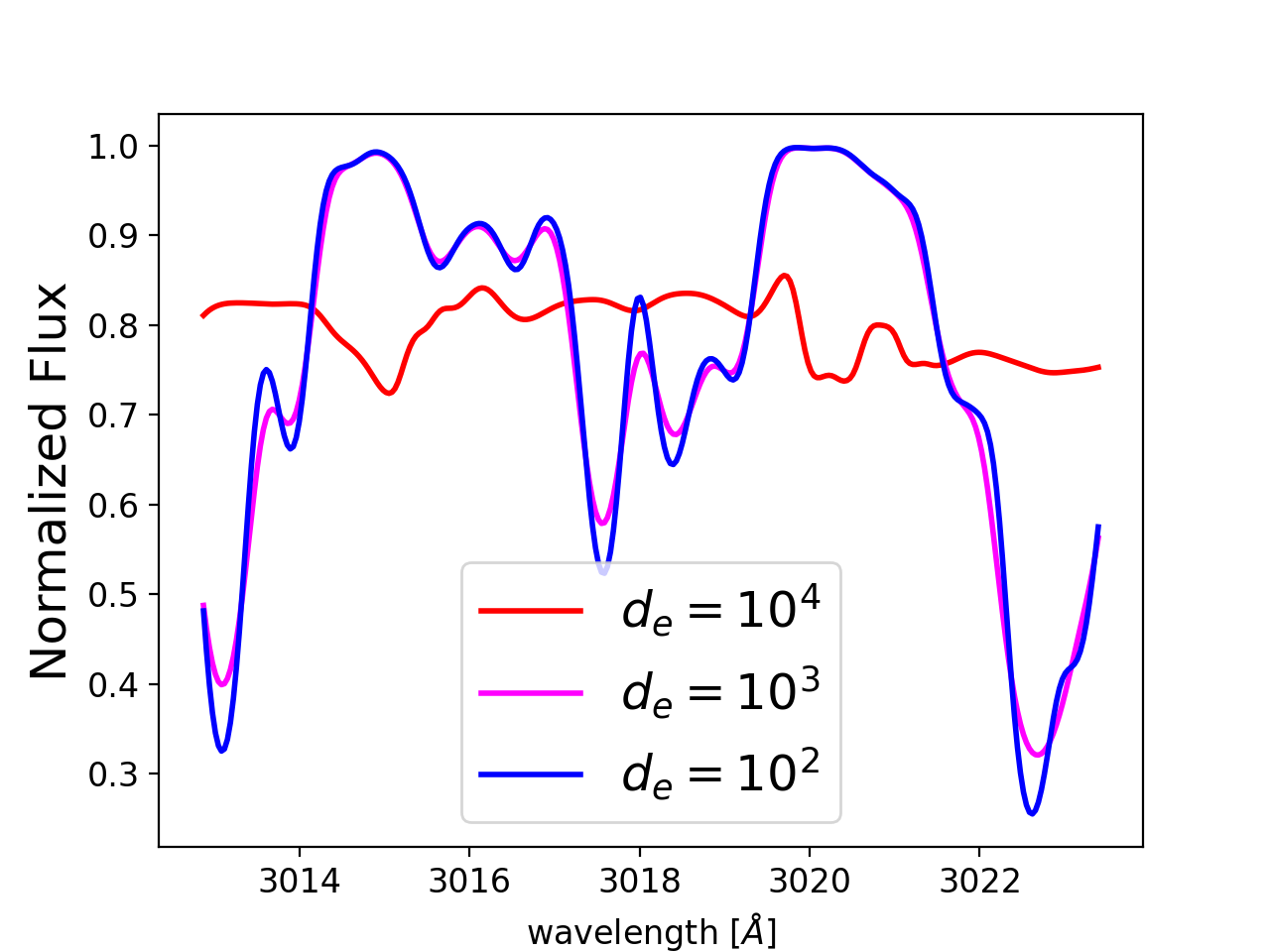}}  
    \subfigure[]{
        \includegraphics[width=0.45\textwidth]{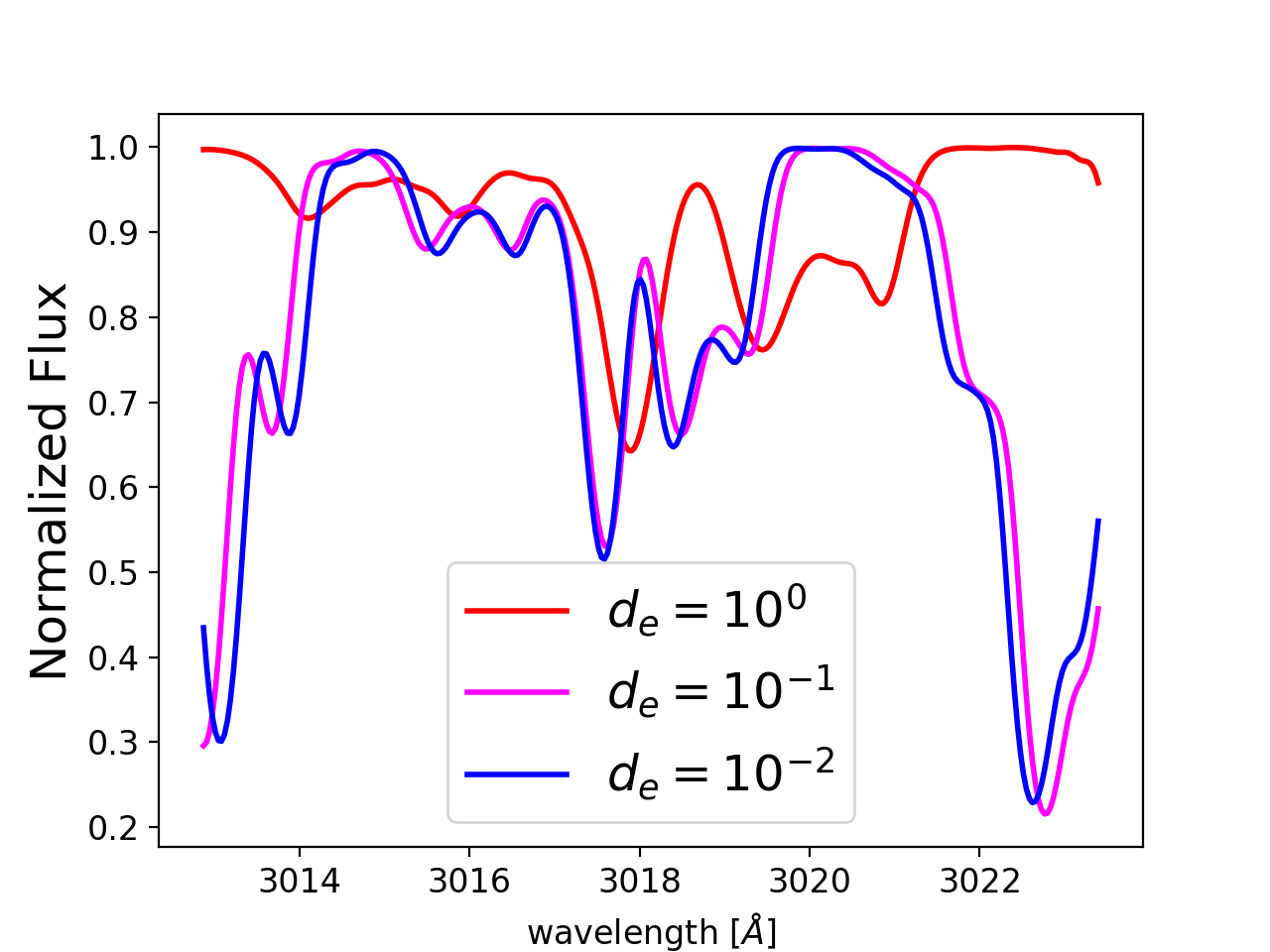}}  
    \caption{Example spectra (without noise) for different masses and couplings, chosen such that same coloured lines display same amplitude of dilaton effect. Panel (a): $m_\phi=10^{-26}\,\mathrm{eV}$ displays line broadening. Panel (b): $m_\phi=10^{-30}\,\mathrm{eV}$ displays line shifting. In the  case of shifting, much stronger effects are visible at lower dilaton coupling.}
    \label{fig: spectra}
\end{figure*}

We simulated mock spectra without coupling to dilatons and with five different couplings in the parameter space of interest. We show some example spectra for different masses and various couplings in Fig.~\ref{fig: spectra}. In the left panel we observe the broadening effect of the dilaton coupling. If the broadening gets too large, there is no absorption structure visible in the spectrum anymore (blue line). Large values of $d$ lead to a ``washing out" of structure, i.e. the broadening due to the coupling to the dilaton precludes the formation of narrow absorption lines at all. In the right panel we show the spectrum from a smaller mass $m_\phi = 10^{-30}\,\mathrm{eV}$. The shifting effect is clearly visible by eye (yellow line versus green line). Again, if the dilaton coupling is too large, the formation of narrow absorption lines is suppressed.

\subsection{\label{ssec: likelihood} Likelihood analysis}

As mentioned earlier, large effects can be seen visually for sufficiently large coupling. But more careful bounds can be derived for profile broadening as small as a pixel size or smaller. We follow here the methodology of the profile likelihood ratio statistic, commonly used for exclusion and discovery limits of dark matter \cite{Cowan:2010js}.

The likelihood ratio statistic compares two hypotheses, $\mathcal{H}_1$ and $\mathcal{H}_0$, where in the former we assume no dilaton field couplings, and in the latter we assume one of the dilaton field couplings as non zero. We further assume experimental/mock data where no signal is present. The goal is to reject the $\mathcal{H}_0$ hypothesis with $90\%$ confidence. We start by writing the likelihood ratio:
\begin{equation}
    \Lambda = \frac{\mathcal{L}(\mathcal{H}_0)}{\mathcal{L}(\mathcal{H}_1)}
\end{equation}
where the $\mathcal{H}_i$ signify we enter the hypothesis parameters in the calculation of the likelihood. The value of the dilaton coupling will be the sole difference between the two. We now define the associated test statistic \cite{Cowan:2010js}:
\begin{equation}
    q=
    \begin{cases}
    -2\log(\Lambda) \text{ if } d_i\geq 0  \\
    0 \text{ otherwise}
    \end{cases}
\end{equation}
From Wilks' theorem \cite{Cowan:2010js} we know that in the limit of many observations per bin (photons per pixels) this statistic converges to a $\frac{1}{2}\chi^2_p$ distribution where p is the number of parameters we are testing (in this case one). The ``significance" of the signal is $\sqrt{q}$. We can compute the likelihood of $\mathcal{H}_i$ from the modified Voigt profile, using:
\begin{equation}
    \mathcal{L}(\mathcal{H}_i)=\prod_{j\in \text{pixels}}\mathcal{G}(F(z_j),F_R(z_j,d_i),\sigma_j)
\end{equation}
where $\sigma_j=\sqrt{\left(F(z_j)/\text{SNR}\right)^2 + \sigma_0^2}$ and $\sigma_0$ dominates the thermal noise distribution at very small fluxes. The normalized flux here can be either real or simulated data.

For real data we first need to determine the HI overdensities, assuming it is a realisation of a cosmological power spectrum including a fraction of dilatons with known cosmological parameters (computed with axionCAMB), see our discussion in Sec. \ref{ssec: multimessenger} \footnote{We make the same assumption when using mock data.}. This yields a reconstructed (inverted \cite{Mueller2020}) LOS flux $F_R$ and a recovered HI overdensity. The recovered LOS assumes a fixed cosmology, but with $d_i=0$. If there is no realization of HI overdensities that fit the artificial power spectrum and the Ly$\alpha$ forest simultaneously without $d_i \neq 0$, we would have a detection. In the non-detection case, we try to find constraints, i.e. we try to find the maximum size of the dilaton coupling such that the fitted HI overdensities are still compatible with the synthetic power spectrum. In essence the recovered HI density profile acts as a background in the search for beyond standard model physics, and is used as input for the $\mathcal{H}_1$ hypothesis. We then use it to calculate a LOS for some $d_i>0$, i.e. input for $\mathcal{H}_0$. As discussed in Sec. \ref{ssec: multimessenger} we assume that the instrumental noise in the spectra dominates over the uncertainties in the power spectrum.

The recovered flux is inherently noisy due to instrumental noise. Therefore, each pixel is modelled as an independent Gaussian $\mathcal{G}$ centred around the recovered flux $F_R(z_j,d_i)$ and with noise $\sigma_j$ where $F(z_j)$ and SNR are the telescope's measured normalized flux and SNR on the LOS (it is $\propto \sqrt{\text{exposure time}}$ and we take the median SNR for simplicity).

For $m_\phi\lesssim 10^{-22}$~eV, the pixel crossing time is shorter than the dilaton coherence time, and we no longer integrate over the Rayleigh distribution. Rather the amplitude of the field is fixed during the crossing time, with probability described by the Rayleigh distribution. We therefore draw a sample from the Rayleigh distribution every coherence interval, see our discussion in Sec. \ref{ssec: mod-voigt-profile}. For the very smallest masses, this poses a statistical problem when the number of coherence intervals drops and the sample from the Rayleigh distribution is not representative for the distribution anymore. However, as we will observe how the likelihoods go with the amplitude (i.e. $q \propto \phi_r^2$), we can marginalize the likelihood over the Rayleigh distribution analytically such that the likelihoods \textit{with} signal are now the weighted average. However, we have to mention that there is a small probability that the true dilaton field is in a state which would give no limits at all, i.e. that it is drawn from the smallest possible values of the Rayleigh distribution (c.f. direct detection limits~\cite{Centers:2019dyn}).

Finally, for future telescopes such as SKA, we can simulate an Asimov data set \cite{Cowan:2010js} to set the flux in each pixel to the expected observations of the flux, which is correct in the limit of many observations (i.e. photons) per bin.

\section{\label{sec: results} Results}

\subsection{Ly$\alpha$}

\begin{figure*}[t!]
\includegraphics[width=1.9\columnwidth]{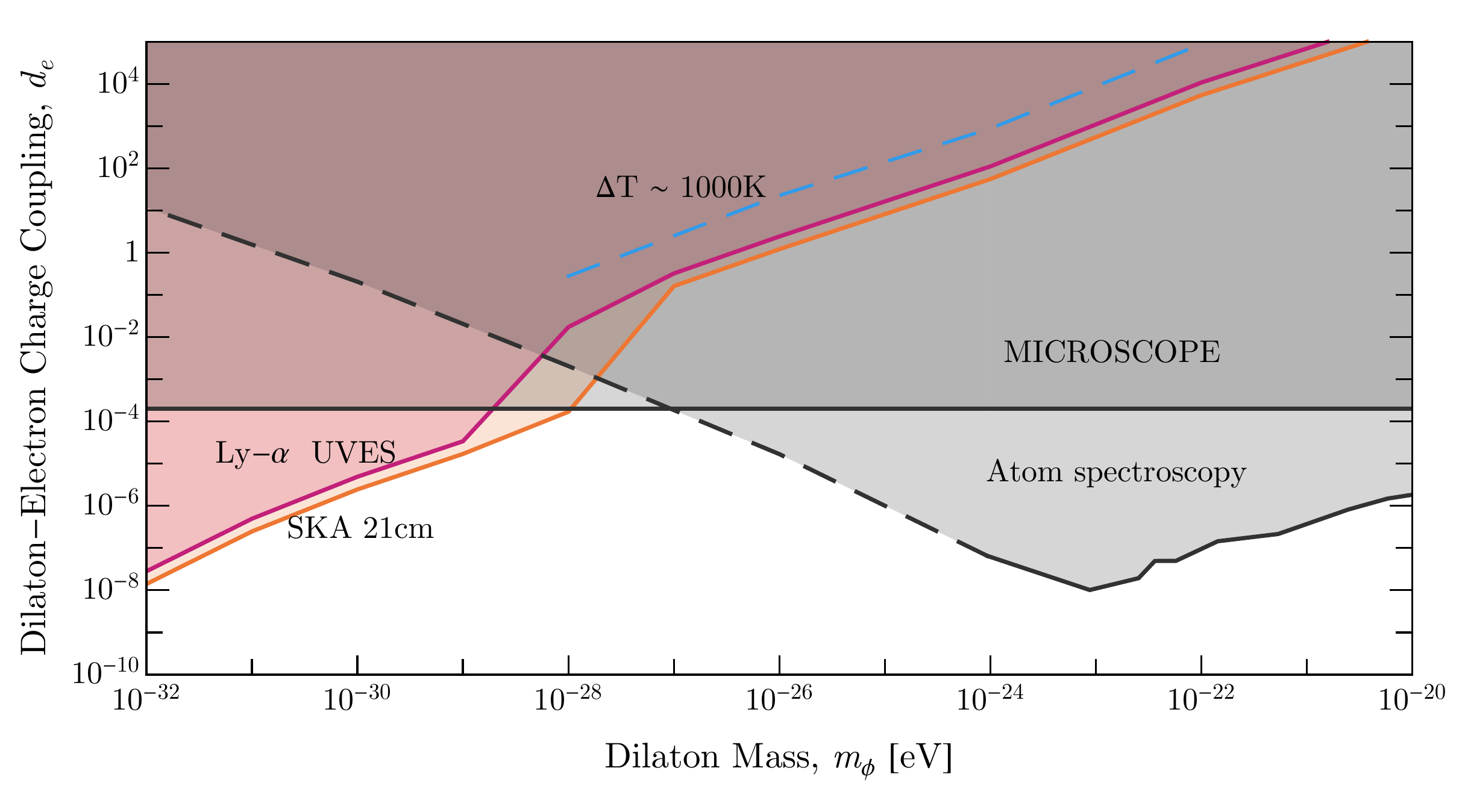}
\caption{\label{fig:de_constraints_all}
Projected and existing constraints on the dilaton coupling $d_e$ as a function of dilaton mass $m_\phi$, in the 90\% confidence limit as in Fig.~\ref{fig:de_constraints}, but extended to larger masses. Moreover, we show direct constraints that we observe from the mismatch between the effective optical depth in our simulation and the observed one (blue).
}
\end{figure*}

Based on the discussion above, we show our forecasts on $d_e$ and $d_m$ in Figs.~\ref{fig:de_constraints} \& \ref{fig:dm_constraints} (details of the forecasting are shown in Appendix \ref{app: fits} and Fig.~\ref{app: fits}). The relation between the two is straightforward: keeping only linear terms in the coupling, the effect is proportional to the power of $\alpha$ or $m_e$ in the Rydberg constant. Therefore bounds on the mass coupling are simply shifted upward by a factor of two compared to the gauge coupling.

For fixed dilaton energy density and masses $t_m<t_{\rm{pix}}$ our bounds are approximately linear as the size of the broadening scales with $d_i \phi_r \propto d_i m_\phi^{-1}$ (and exactly linear in the limit of a zero-variance sample from the Rayleigh distribution). To this we add a factor of $\sqrt{\Omega_{\phi}/\Omega_{DM}}$ such that below $m_\phi=10^{-20}$~eV the constraints do not follow a straight line. The choice of $\Omega_\phi$ constraints from Refs.\cite{Lague2021,Kobayashi:2017jcf} affects our constraints as $\Omega_\phi$. The corresponding factors can be read off Tab.~\ref{tab: implementations}. We also note $q \propto d_i^4$ at these masses ($t_m<t_{\rm{pix}}$), due to the Gaussian likelihoods and the $\phi_m^{-2}$ distribution in the modified Voigt profile.

In the $t_m>t_{\rm{pix}}$ regime ($m_\phi \lesssim 10^{-28}$~eV) we also recover linear bounds before accounting for the DM fraction. However, we obtain slightly improved constraints \footnote{Note that the limits for $m_\phi \lesssim 10^{-29}\,\mathrm{eV}$ are marginalized, we refer the reader to our discussion in Sec. \ref{ssec: likelihood}.}. This is due to a new behaviour, not seen when averaged, of overall displacement of the peaks of the absorption profiles. Peak displacement implies $q \propto d_i^2$ for the smallest masses ($t_\mathrm{pix} > t_m$). This changed scaling occurs because of the loss of the $\phi_m^{-2}$ distribution in the modified Voigt profile.

A quick consistency check confirms the results obtained by our likelihood procedure. In case of broadening the additional broadening of the line is roughly given by the term $\Delta s \sim 2c\kappa d_i \phi_r \sim 0.2\,\mathrm{km}/\mathrm{s}$ for the bounds obtained here. This corresponds to roughly $100\,\mathrm{K}$ of thermal Doppler broadening, a precision well achievable once the polytropic index is fixed and peculiar velocities are known \citep{Mueller2021}. Various estimation procedures for the temperature seem to converge to a common mean temperature of roughly $10^4\,\mathrm{K}$, but only to a uncertainty of roughly $1000\,\mathrm{K}$ which is due to astrophysical uncertainties (photoionization rate, polytropic index, peculiar velocities, ...) and uncertainties in the cosmic model \citep{Becker2011, Boera2014, Walther2019, Gaikwad2020}. Following this analogy we present in Fig.~\ref{fig:de_constraints_all} also the couplings at which $\Delta s$ exceeds the current thermal cosmic uncertainty of $1000\,\mathrm{K}$. These bounds are immediately applicable as a broadening effect of this size would have already led to an observed signal in the search estimation of the temperature of the IGM (neglecting degeneracies between thermal modelling and the dilaton effect). However, they do not represent the bounds that are possible once the astrophysical and cosmological models are fixed.


We also show competing bounds in Figs.~\ref{fig:de_constraints} \& \ref{fig:dm_constraints}. Equivalence principle tests \cite{Berge:2017ovy,Wagner:2012ui} are the strongest bounds for the lowest masses. These tests use Eötvos-type torsion pendulum experiments to look for non-oscillating Yukawa-type potentials with scale parameter $m_\phi$ which create deviations from the expected equality between forces felt by two objects of different composition or mass. Such a force would be expected from virtual dilaton fields centred around objects from couplings to the electromagnetic gauge sector or electron mass terms \cite{Wagner:2012ui}. At low masses where $l_{\rm{coh}}>l_{\rm{exp}}\approx 10^{-13}$~eV the range of the Yukawa force is longer than the experiment length and variations in the potential can no longer be detected. In this regime one can detect deviations from the gravitational forces between the pendulum's masses (i.e. in the composition dependent case where they are charged differently under the fifth force). This regime corresponds to the maximum sensitivity 
which no longer varies with $m_\phi$ and covering the whole range of masses in the figure. Other types of bounds require long integration times, e.g. by looking for time deviations (or oscillations) in atomic spectroscopy and atomic clocks \cite{Arvanitaki:2014faa,VanTilburg:2015oza,Hees:2016gop}. The experimental reaches are therefore typically limited below $m_\phi<10^{-24}\text{~eV}\sim \mathcal{O}(1\text{ year}^{-1})$, beyond which the experiment looks for a time gradient in the field (e.g. \cite{VanTilburg:2015oza}, or searches for time oscillating neutron EDM~\cite{Abel:2017rtm}). Although detecting time gradients of course gets harder as the oscillation frequency is slower, it can be compensated by higher couplings, since the size of the effect (i.e. change in electronic/nuclear transition energies) grows approximately as fast with coupling as it decreases with frequency \cite{VanTilburg:2015oza}. We therefore show here linearly extrapolated bounds from \cite{Hees:2016gop}\footnote{Note the non-linear behavior comes from rescaling bounds using dilaton density fractions in Table~\ref{tab: implementations}. Atom spectroscopy bounds (somewhat inconsistent) above $10^{-24}$~eV assume 100\% of DM is dilatons, whereas the MICROSCOPE experiment does not require assumptions on the background dilaton density.}.


\subsection{SKA}

Next we discuss the impact of the SKA telescope \cite{Aharonian:2013av,Furlanetto:2006jb} on our bounds. Using identical arguments to the Ly$\alpha$ line, hydrogen's 21 cm hyperfine transition line can be modelled with similar absorption/emission profiles \cite{Furlanetto:2006jb}. The effect of the dilaton field on the profile is similar to the Ly$\alpha$ line coupling through $(d_e,d_{m_e})$ up to a factor of 2, since the absorption energy is calculated as:
\begin{align}
    \Delta E_{\mathrm{21cm}} = \frac{4}{3}g_eg_p\alpha^2\frac{m_e}{m_p}\mathrm{Ry}
\end{align}
where $\mathrm{Ry}\propto\alpha^2 m_e$ as we saw in \eqref{eq: Lyalpha energy}, $g_e$ and $g_p$ are the gyromagnetic factors of the electron and proton respectively and $m_p$ is the proton mass.

However comparing the Ly$\alpha$ and 21 cm signals this way neglects the more complex astrophysics of the 21cm signal during the epoch of reionization, which requires careful modelling of non-linear physics which are still unknown (e.g. \cite{Choudhury2016,Furlanetto:2006jb}). As more parameters are needed to model the 21cm signal, this also introduces the risk of degeneracies with the dilaton effect. This work assumes that we do not see a dilaton effect, which requires an absence of degeneracies. We will therefore neglect any possible degeneracies, but future work is needed to investigate whether they effect our bounds.

With this caveat, we can extrapolate our results for searches in the Ly$\alpha$ forest to 21 cm searches, and infer new rough bounds by replacing the SNR and number and size of pixels (called ``voxels" here). The number of voxels of an interferometric 21 cm radio survey are calculated using the angular and spectroscopic resolution of the primary beam of an SKA I LOW survey, assumed to be $\Delta\theta^2=\lambda_{\mathrm{21cm}}^2/D_{\mathrm{dish}}^2=0.5$~deg$^2$ and $\Delta\nu=10^4$~Hz over a survey area $S_{\mathrm{area}}=$25000~deg$^2$ and the 80-200~MHz
band \cite{Santos:2015gra}. The SNR is calculated using the radiometer equation for interferometers \cite{Santos:2015gra} which for resolutions of order the primary beam size $\Delta\theta^2$ simplifies to:
\begin{align}
    \mathrm{SNR}=\frac{\delta T_{21cm}}{T_{\mathrm{sys}}}\sqrt{\Delta\nu t_p N_b}
\end{align}

Here we assume $t_{\mathrm{tot}}$=12.500 hours total integration time, thus pointing time $t_p=t_{\mathrm{tot}}\frac{\Delta\theta^2}{S_{\mathrm{area}}}=$ 15 minutes. For simplicity we assume an average $T_{\mathrm{sys}}=1100$~K (where we used $T_{\mathrm{sys}}=250(z/7)^{2.75}$~K \cite{Braun:2019gdo})
and $\delta T_{21cm}\approx 10mK$ variations of $T_{\mathrm{21cm}}$ along the line of sight at $z\sim 13$. We also assume a relatively constant (or average) SNR across the frequency band, ie we assume $\delta T_{21cm}$ and $T_{\mathrm{sys}}$ follow a similar power law in $z$ down to $z\approx 8$ where we assume reionization ends \cite{Choudhury2016,PhysRevD.100.063538,PhysRevD.106.043529,Furlanetto:2006jb}. $N_b$ is the number of nearest-neighbor baselines used near the largest resolution angle $\Delta\theta^2$, which we estimate conservatively as $N_b\approx N_d$, where $N_d=257$ is the number of dishes for SKA I MID. As mentioned in \cite{Santos:2015gra} we assume the primary beam size to be the largest angle at which we can build single images. We can thus obtain an SNR $\approx 0.6$, which we use for bounds in Figs.~\ref{fig:de_constraints}, \ref{fig:dm_constraints} \& \ref{fig:de_constraints_all}. Since (analogously to Ly-$\alpha$) $q\propto d_i^2$ at low dilaton masses, where now $t_{pix}\approx 10h^{-1}$kpc for SKA I, and $q\propto d_i^4$ at intermediate/high mass, and since $q\propto$ (\# voxels)SNR$^2$, bounds on both dilaton couplings go as (\# voxels)$^{1/4}$SNR$^{1/2}$ at intermediate/high mass and (\# voxels)$^{1/2}$SNR at low mass. We also account for the linear effect of  $\langle\phi_r\rangle=\sqrt{2 \bar{\rho}_\phi(z)}/m_\phi \propto(1+z)^{3/2}$ on the bounds.

Using SKA with large beam angle optimizes our bounds as using smaller angles $\Delta\theta^2$ would increase the number of voxels by $\Delta\theta^{-2}$, but would decrease the SNR by a factor of $D_{\mathrm{dish}}^2/D_b^2\propto\Delta\theta^2$ because the baseline length $D_b$ enters the definition of that angle. Varying the spectral resolution on the other hand does not affect our bounds as SNR and voxel effects cancel. Nevertheless we choose smaller resolution as this places us in the preferred visibility "window" of 21cm foreground removal (see e.g. \cite{Gagnon-Hartman:2021erd} and refs therein) and optimizes $\delta T_\textrm{21cm}$ \cite{Furlanetto:2006jb,PhysRevD.100.063538}.


Finally, looking ahead, SKA phase II is predicted to have even better SNR ($\times$ 10) \cite{Braun:2019gdo} and similar (or slightly larger) pixel number and could further improve our projections in Figs.~\ref{fig:de_constraints} \& \ref{fig:dm_constraints} by a factor of $10^1-10^2$.

\section{\label{sec: discussion} Discussion}

The possible existence of dilatons and other light scalar field DM is well motivated by appeal to string theory. Via coupling to the fine structure constant and electron mass, dilaton DM can lead to observable effects in laboratory and astronomical observations. In an era of rapidly increasing volume of astronomical data, it would be unwise not to fully exploit this for these signs of couplings to beyond the standard model physics. Many tools such as \textsc{REGLYMAN} and \textsc{axionCAMB}, which we use here, have been developed to predict and analyze astrophysical data, and can be adapted to compute dilaton effects.

The dilaton couplings we have considered lead to shifts in the energy of atomic transitions in Hydrogen, the size of which are correlated with the dilaton DM density. We have introduced a model for the observable effect this energy shift leads to in spectroscopic surveys such as UVES SQUAD or SKA. Convolving the atomic energy level shift with the distribution of the classical dilaton DM field predicted from cosmic structure formation with quasar emission and absorption spectra our model predicts that dilaton DM induces a line broadening at masses above the spectra's pixel crossing frequencies, and a line shift at lower masses. Assuming a fixed cosmology, we fit for neutral hydrogen overdensities and identify these effects in simulated data. By considering the signal to noise of these effects in our simulated data, we are able to forecast the ability of UVES SQUAD and SKA to constrain dilaton couplings in the future. We find that laboratory constraints can be surpassed by several orders of magnitude, if dilatons make up just a small fraction of the total cosmic DM.

A follow-up study may enhance this by cross-correlating the peak displacement/broadening with alternative measurements that scale differently with the dilaton coupling (i.e. standard candles, 21 cm signals, Ly$\beta$ forest). We further find that the line shifting regime is a more statistically powerful effect than broadening, yielding larger projective constraints on the couplings. Up to a dependence on $\Omega_\phi/\Omega_d$, the bounds go linearly with the mass in each regime.


Our results must of course take into account the presence of uncertainties, degeneracies and model dependence in the underlying assumptions (e.g. cosmological parameters, instrumental response patterns, temperature modelling, and systematic uncertainties introduced by the deconvolution process), which we don't account for in our nuisance parameters as we assume noise from Ly$\alpha$ survey data to dominate. Further study of the statistical interplay of these effects and the addition of alternative measurements (standard candles, 21 cm signals,Ly$\beta$ forest) may further improve the quality of the bounds in our study. In the past, uncertainties in the thermal evolution of the IGM limited the predictive power of Ly$\alpha$ forest surveys. This does not limit our analysis here: the dominating effect for ultralight dilatons was found to be a displacement of the lines, while thermal inference mainly affects the width of the absorption features. Peculiar velocities and redshift space distortions may also affect the observables at play, although recent developments in estimating these distortions from three dimensional density maps constructed from closely neighbouring Ly$\alpha$ lines may alleviate this problem. 

Our study here harnesses a unique power of astronomical surveys: cosmological integration times. Only astronomical surveys can probe the time varying effect of oscillating fields with masses much below $10^{-24}$~eV, thus possibly putting new constraints on dilatons at the smallest mass scales. At higher masses we mention that these bounds, although not competitive at masses $m_\phi \gtrsim 10^{-28}$~eV, also extend into to the mass range accessible for atom spectroscopy. Furthermore the expected improvement in decades to come in SNRs and increases in the amount of high SNR data (e.g. \cite{SimonsObservatory:2018koc,Gebhardt:2021vfo,Bull:2014rha}), may further improve the ability of cosmological data to constrain dilaton couplings. 21 cm surveys are a new frontier in observational cosmology, and the SKA phase I survey is expected to be collecting data before the end of the decade. Further the proposed improvements of an SKA phase II survey could improve the bounds shown here (we use parameters consistent with SKA phase I)\cite{Braun:2019gdo}. Meanwhile, combining data from UVES SQUAD and near future surveys such HETDEX \cite{Gebhardt:2021vfo} may improve our constraints on dilaton couplings. The novel ELT/HIRES \cite{2021Msngr.182...27M} instrument might also play a crucial role in further pushing these bounds due to its significantly increased sensitivity and SNR of high resolution spectra ($R \sim 100000$) compared to VLT/UVES observations. We also mention future cosmological parameter fits using joint \emph{Planck} and Simons Observatory/CMB-S4 survey data would allow to directly observe/constrain ultralight dilaton fields to sub-percent levels of the dark matter density~\cite{Hlozek:2016lzm,CMB-S4:2016ple,SimonsObservatory:2018koc}, a key ingredient necessary to discover dilatons via the methods discussed in this paper.


Finally, there is much space still for future work putting Ly$\alpha$ to work to constrain new physics. First, we omit to constrain the coupling of the 21 cm line to the inverse proton mass, $m_p^{-1}$, which is affected by the dilaton's coupling to the QCD mass scale and the quark masses. However, as discussed in \cite{Damour:2010rp,Hees:2016gop} the dependence is complex and we do not determine the constraints here. Analyses of Ly$\alpha$ data can also potentially reveal or constrain many other effects, such as chameleon dark energy or isocurvature modes from the early universe \cite{Ballesteros:2021bee,Brax:2004px}. We hope the present study encourages further work in these directions.

\section*{Data Availability}

We will make our software and mock data publicly available as part of the second release of the \textsc{REGLYMAN} software \cite{reglyman} upon acceptance and reasonable request.

\acknowledgments 

LH and HM have contributed equally to this work. LH would like to thank Bassem Alachkar, Philip Bull, Jonathan Pober, Michael Wilensky and Jacob Burba for discussions on 21 cm astronomy, Jurek Bauer for discussion on axion structure formation as well as Alan Roy, Derek Jackson Kimball, Christopher McCabe, and John Carlton for helpful discussions and reading early versions of the manuscript. LH is supported by the King's Cromwell Scholarship. HM received financial support for this research from the International Max Planck Research School (IMPRS) for Astronomy and Astrophysics at the Universities of Bonn and Cologne. DJEM is supported by an STFC Ernest Rutherford Fellowship.

\bibliography{main}

\appendix{}

\section{IGM Physics} \label{sec: igm-physics}
The temperature of the IGM is modelled by a power law \cite{Hui1997, Mueller2021}:
\begin{align}
    T(x, z) = T_0(z) \Delta^{\zeta-1}, \label{eq: temperature}
\end{align}
where $\Delta = \rho_b / \langle \rho_b \rangle$ is the fractional baryonic density perturbation and $T_0$ the temperature at mean density (i.e. note $T_0$ is not the mean temperature). $T_0$ was measured recently at redshifts $z \sim 2.5$ with a wide range of different methods, e.g. by \cite{Becker2011, Garzilli2012, Boera2014, Hiss2018, Telikova2019, Walther2019, Gaikwad2020, Mueller2021}. While there is significant scatter, the different estimates coincide at a temperature between $10000\,\mathrm{K}$ and $15000\mathrm{K}$ for redshifts $z \sim 2-3$. Similar work was done for the power-law index $\zeta$, e.g. see \cite{Garzilli2012, Bolton2014, Hiss2018, Rorai2018, Telikova2019, Walther2019, Gaikwad2020, Mueller2021} predicting indices around $\zeta = 1.4$ in the redshift range of interest. Under the assumption that the IGM is mainly composed of neutral hydrogen, the baryonic density perturbation can be related to the neutral hydrogen density \cite{Hui1997, Nusser1999, Gallerani2006}:
\begin{align}
    n_\mathrm{HI} (x, z) = \hat{n}_\mathrm{HI}(z) \Delta^\alpha(x, z), \label{eq: density}
\end{align}
where $\alpha = 2.7 - 0.7 \zeta$ and $n_\mathrm{HI}$ is the neutral hydrogen density at mean baryonic density distribution which depends in particular on the photoionization rate of the IGM \cite{Choudhury2001, Mueller2021}.

\section{Fits} \label{app: fits}
\begin{figure*}
    \centering
    \subfigure[$m_\phi = 10^{-20}\,\mathrm{eV}$]{
        \includegraphics[width=0.3\textwidth]{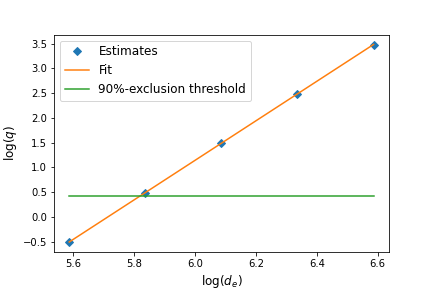}}
    \subfigure[$m_\phi = 10^{-22}\,\mathrm{eV}$]{
        \includegraphics[width=0.3\textwidth]{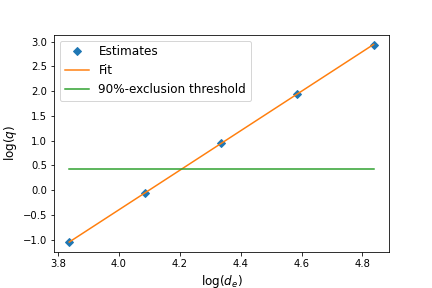}}
    \subfigure[$m_\phi = 10^{-24}\,\mathrm{eV}$]{
        \includegraphics[width=0.3\textwidth]{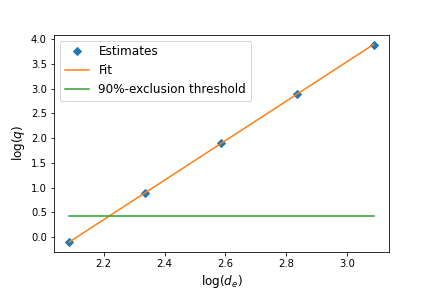}}
    \subfigure[$m_\phi = 10^{-26}\,\mathrm{eV}$]{
        \includegraphics[width=0.3\textwidth]{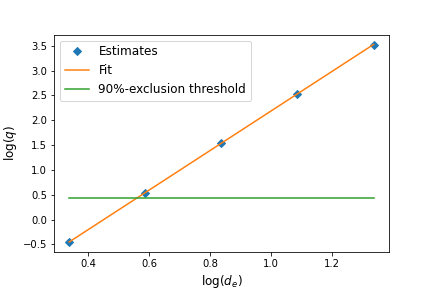}}
    \subfigure[$m_\phi = 10^{-27}\,\mathrm{eV}$]{
        \includegraphics[width=0.3\textwidth]{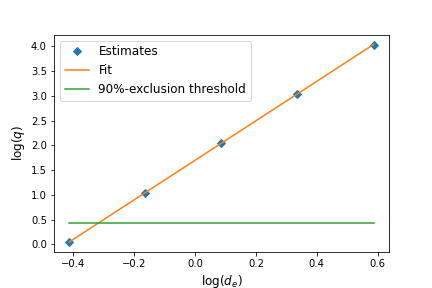}}
    \subfigure[$m_\phi = 10^{-28}\,\mathrm{eV}$]{
        \includegraphics[width=0.3\textwidth]{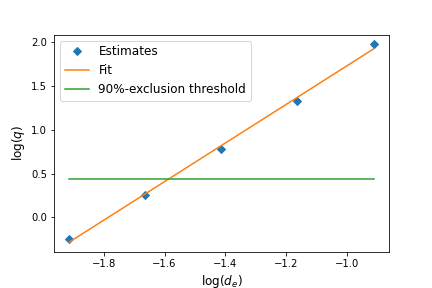}}
    \subfigure[$m_\phi = 10^{-29}\,\mathrm{eV}$]{
        \includegraphics[width=0.3\textwidth]{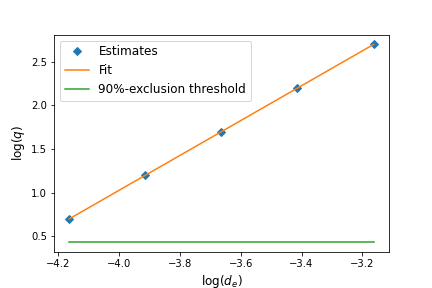}}
    \subfigure[$m_\phi = 10^{-30}\,\mathrm{eV}$]{
        \includegraphics[width=0.3\textwidth]{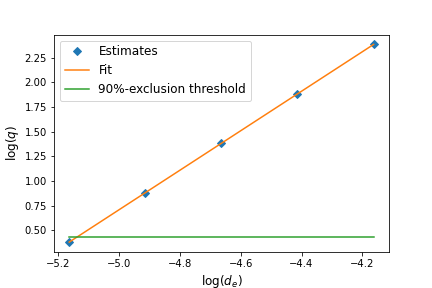}}
    \subfigure[$m_\phi = 10^{-31}\,\mathrm{eV}$]{
        \includegraphics[width=0.3\textwidth]{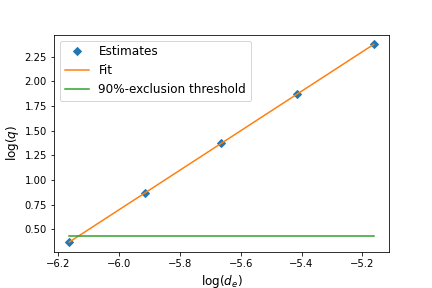}}
    \subfigure[$m_\phi = 10^{-32}\,\mathrm{eV}$]{
        \includegraphics[width=0.3\textwidth]{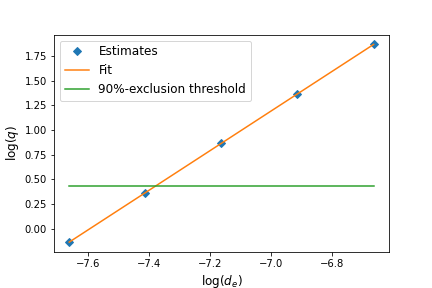}}
    \caption{Computed likelihoods (blue diamonds), i.e. $\log(q)$ for varying coupling constants $d_e$, with linear fit in double logarithmic space (orange) and $90\%$-exclusion threshold.}
    \label{fig: likelihood_fits}
\end{figure*}

For every mass, we computed the spectra without coupling to dilatons and with five different couplings that are unequal to zero. We compute the likelihoods for all these spectra with our likelihood analysis presented in Sec. \ref{ssec: likelihood}. Then, for a fixed mass, we fit these likelihoods as a function of coupling linearly in double-logarithmic space. We compute our constraints by the corresponding fit values for the $90\%$ likelihood threshold ($q=2.71$). Our fits are presented in Fig.~\ref{fig: likelihood_fits}.

\end{document}